\documentclass[10pt,conference]{IEEEtran}
\usepackage[utf8]{inputenc}

\usepackage{framed}

\usepackage{amsmath,amsfonts,amsthm, amssymb, array,url}

\usepackage{hyperref}
\usepackage{booktabs}

\newcommand{\PreserveBackslash}[1]{\let\temp=\\#1\let\\=\temp}
\newcolumntype{C}[1]{>{\PreserveBackslash\centering}p{#1}}
\newcolumntype{R}[1]{>{\PreserveBackslash\raggedleft}p{#1}}
\newcolumntype{L}[1]{>{\PreserveBackslash\raggedright}p{#1}}

\usepackage[linesnumbered,ruled,vlined]{algorithm2e}

\SetAlFnt{\small}
\SetAlCapFnt{\small}
\SetAlCapNameFnt{\small}

\usepackage[edges]{forest}

\usepackage{multirow}

\usepackage{balance}
\usepackage{comment}
\usepackage{ifthen}
\usepackage{subcaption}
\usepackage[utf8]{inputenc}
\usepackage[T1]{fontenc}

\SetCommentSty{mycommfont}

\AtBeginDocument{%
  \providecommand\BibTeX{{%
    \normalfont B\kern-0.5em{\scshape i\kern-0.25em b}\kern-0.8em\TeX}}}

\emergencystretch=\maxdimen

\usepackage{listings}

\definecolor{codegreen}{rgb}{0,0.6,0}
\definecolor{codegray}{rgb}{0.5,0.5,0.5}
\definecolor{codepurple}{rgb}{0.58,0,0.82}
\definecolor{backcolour}{rgb}{0.95,0.95,0.92}

\newboolean{showcomments}
\setboolean{showcomments}{true}
\ifthenelse{\boolean{showcomments}}
{
	\definecolor{myyellow}{RGB}{255, 228, 26}
	\definecolor{myblue}{RGB}{50, 50, 220}
	\newcommand{\nb}[2]{
		{\sf
			\fcolorbox{myyellow}{yellow}{\scriptsize\textbf{#1}}%
			$\blacktriangleright$%
			{\color{myblue}\fontsize{7pt}{8pt}\selectfont\textbf{#2}}%
		}%
	}
}
{
	\newcommand{\nb}[2]{}
}

\newcommand{\ali}[1]{\nb{Ali}{#1}}

\newcommand{\header}[1]{\par\smallskip\noindent\textbf{#1.}}

\newcommand{\apiCall}[1]{$\mathcal{A}_{#1}$}
\newcommand{\request}[1]{$\mathcal{R}q_{#1}$}
\newcommand{\response}[1]{$\mathcal{R}s_{#1}$}
\newcommand{\pathNode}[1]{$\nu_{#1}$}
\newcommand{\apiGraph}[1]{$\mathcal{G}_{#1}$}
\newcommand{\graphPath}[1]{$\zeta_{#1}$}

\newcommand{\q}[1]{RQ\textsubscript{#1}}

\newcommand{\toolname}{\textsc{ApiCarv}\xspace}

\theoremstyle{definition}
\newtheorem{definition}{Definition}

\newcommand{\code}[1]{\texttt{\fontsize{8.5}{9}\selectfont #1}}

\usepackage{color}
\usepackage{textcomp}
\definecolor{listinggray}{gray}{0.9}
\definecolor{lbcolor}{rgb}{0.9,0.9,0.9}

\usepackage{enumitem}

\newlist{researchquestions}{enumerate}{1}
\setlist[researchquestions]{label*=\textbf{RQ\arabic*},  itemindent=1em}

\usepackage[justification=centering]{caption}

\usepackage[]{xcolor}
\usepackage{colortbl}

\usepackage{soul}
\usepackage{tcolorbox}

\newcommand{\finding}[1]{
\begin{tcolorbox}[boxsep=0pt,left=7pt,right=7pt,top=5pt,bottom=5pt]
{#1}
\end{tcolorbox}
}

\begin{document}

\title{
Carving UI Tests to  Generate API Tests and \\ API Specification
} 

\author{
\IEEEauthorblockN{Rahulkrishna Yandrapally}
\IEEEauthorblockA{University of British Columbia\\
  Vancouver, BC, Canada \\
rahulyk@ece.ubc.ca}
\and
\IEEEauthorblockN{Saurabh Sinha}
\IEEEauthorblockA{IBM Research \\
  Yorktown Heights, NY, USA \\
sinhas@us.ibm.com}\\
\and
\IEEEauthorblockN{Rachel Tzoref-Brill}
\IEEEauthorblockA{IBM Research\\
  Haifa, Israel \\
rachelt@il.ibm.com}\\
\and
\IEEEauthorblockN{Ali Mesbah}
\IEEEauthorblockA{University of British Columbia\\
  Vancouver, BC, Canada \\
amesbah@ece.ubc.ca}\\
}

\maketitle

\thispagestyle{plain}
\pagestyle{plain}

\newlength{\textfloatsepsave}
\setlength{\textfloatsepsave}{\textfloatsep}
\setlength{\textfloatsep}{3pt}
\setlength{\floatsep}{3pt}

\begin{abstract}

Modern web applications make extensive use of API calls to update the UI state
in response to user events or server-side changes. For such applications,
API-level testing can play an important role, in-between unit-level testing and
UI-level (or end-to-end) testing. Existing API testing tools require API
specifications (e.g., OpenAPI), which often may not be available or, when
available, be inconsistent with the API implementation, thus limiting the
applicability of automated API testing to web applications.
In this paper, we present an approach that leverages UI testing to enable
API-level testing for web applications. Our technique navigates the web
application under test and automatically generates an API-level test suite,
along with an OpenAPI specification that describes the application's server-side
APIs (for REST-based web applications). A key element of our solution is a
dynamic approach for inferring API endpoints with path parameters via UI
navigation and directed API probing.  We evaluated the technique for its
accuracy in inferring API specifications and the effectiveness of the ``carved''
API tests.  Our results on seven open-source web applications show that the
technique achieves 98\% precision and 56\% recall in inferring endpoints. The
carved API tests, when added to test suites generated by two automated REST API
testing tools, increase statement coverage by 52\% and 29\% and branch coverage
by 99\% and 75\%, on average.
The main benefits of our technique are: (1) it enables API-level testing of web
applications in cases where existing API testing tools are inapplicable
and (2) it creates API-level test suites that cover server-side code
efficiently while exercising APIs as they would be invoked from an
application's web UI, and that can augment existing API test suites.

\end{abstract}



\section{Introduction}

Software applications routinely use web APIs for establishing
client-server communication. In particular, they increasingly
rely on web APIs that follow the REST (REpresentational State Transfer)
architectural style~\cite{fielding2000architectural} and are referred to as
RESTful or REST APIs. A typical REST API call starts with an HTTP request made by the client, e.g., the front-end of a web application running in the browser, and ends
with a response sent by the server or the back-end of the application. To help
clients understand the operations available in a service and the request and
response structure, REST APIs are often described using a specification
language, such as OpenAPI~\cite{openapi}, API
Blueprint~\cite{apiblueprint}, and RAML~\cite{raml}.

Web application testing is typically performed at multiple levels, each employing different techniques/tools and with different end goals. Unit-level testing of client- and server-side components focuses on validating the low-level algorithmic and implementation details and achieving high code coverage. In contrast, UI-level testing (also called end-to-end testing) focuses on covering navigation flows from the application's web UI, exercising various tiers of the application in end-to-end manner.
In between unit and UI testing, API testing
places the focus of testing on the operations of a service as well as sequences
of operations; it exercises the server-side flows more comprehensively than unit
testing but without going through the UI layer. API-level testing is guided by code-coverage goals as well as API-coverage goals
(e.g.,~\cite{martinlopez2019criteria}).

For web applications that use RESTful APIs whose specifications are available, a number of automated testing techniques and
tools (e.g.,~\cite{arcuri2019restful,
  atlidakis2019restler, viglianisi2020resttestgen, karlsson2020quickrest,
  martin2020restest, godefroid2020intelligent, Zac2021schemathesis,
  corradiniautomated, segura2017metamorphic, laranjeiro2021black,
  stallenberg2021improving, wu2022combinatorial, liu2022morest}) could be
leveraged for API-level testing. These tools take as input an API specification,
and automatically generate test cases for exercising API endpoints defined in
the specification.  However, in practical scenarios, using these tools may not
always be possible.

First, for applications that do not have RESTful APIs, such tools are
inapplicable. This rules out large classes of web applications, such as Java
Enterprise Edition
as well as legacy web applications, which could benefit just as well from
automated API-level testing. Second, for web applications with RESTful APIs, API
specifications may not be available. This can occur because of different
reasons, often because the APIs are meant for use by the specific web
application only or applications within an enterprise, and not exposed for
invocation by external clients. Thus, formal API documentation is considered
less important and not done due to development pressures or other factors.
Moreover, even when API specifications are available, they can be obsolete and
inconsistent with API implementations~\cite{10.1145/3491038}. As a web application and its APIs
evolve, the specifications---which can be large and complex---often fail to
co-evolve due to the maintenance effort involved.\footnote{Although there exist tools for automatically documenting REST APIs
(e.g., SpringFox~\cite{springfox} and SpringDoc~\cite{springdoc}), which can
reduce the cost of keeping API specifications up-to-date with API
implementations, their applicability is limited (e.g., to web applications
implemented using Spring Boot~\cite{springboot}).}


In this work, we address the challenges of enabling automated API-level test generation
universally for web applications, irrespective of whether they use RESTful web
services, and automatically inferring OpenAPI specifications for web applications
that use RESTful APIs. We present a dynamic technique that executes the web
application via its UI to automatically create (1) API-level test cases that
invoke the application's APIs directly, and (2) a specification describing the
application's APIs that can be leveraged for development and testing purposes.

Although prior work has investigated carving unit-level tests from system-level executions using code-instrumentation techniques (e.g.,~\cite{DBLP:journals/tse/ElbaumCDJ09, ICSM-2007-JoshiO, DBLP:conf/sigsoft/XuRTQ07}), no technique exists for carving API-level test cases from UI paths or test cases.

Our technique monitors the network traffic between the browser and the server,
while navigating the application's UI, and records the observed HTTP requests
and responses. Then, it applies filtering to exclude the requests (and their
responses) that are considered unnecessary for API testing. Next, it builds an
API graph from the filtered requests and analyzes the graph to infer a
specification that captures API endpoints (or resource paths), the applicable
HTTP methods (e.g., \textsc{get}, \textsc{post}), and the request/response
structure for each API operation (combination of HTTP method and API endpoint).
A key feature of our technique is that it infers path parameters or variables
for API endpoints from concrete endpoint instances observed during the navigation of
UI paths. Moreover, it uses a novel algorithm for directed API probing and API
graph expansion to discover more concrete endpoint instances that would
otherwise be missed by UI path navigation alone.

The generated API specification can serve as documentation for server-side APIs
of a web application (even if the APIs are not RESTful) and also be fed as input
into an existing API testing tool for automated test generation
(e.g.~\cite{arcuri2019restful, atlidakis2019restler, viglianisi2020resttestgen,
  martin2020restest, Zac2021schemathesis}) or used for checking inconsistencies
in existing API specifications (for RESTful APIs).


The ``carved'' API test cases are derived from UI paths.  Because these tests bypass the UI layer, they execute much more efficiently and are less prone to brittleness than UI-level tests. Yet, they cover the same server-side code as the UI paths from which they are derived and exercise the APIs in ways that they would be invoked from the UI. 

We implemented our technique in a tool called \toolname{} that takes as input 
a UI test suite (generated or manually written) and carves API test cases and an OpenAPI specification.


We conducted an empirical study on seven open-source web applications to
evaluate the technique's effectiveness in carving API tests and inferring
OpenAPI specifications.  With respect to test carving, our results are
two-fold. First, they quantify the expected benefits of carved API tests: the
tests attain similar coverage as the UI test paths from which they are derived,
but at a fraction of the execution cost of UI tests: on average, more than 10x
reduction in test execution time.  Second, our results illustrate that carved
API tests can increase the coverage achieved by two automated API test
generators, EvoMaster~\cite{evomaster, arcuri2019restful} and
Schemathesis~\cite{schemathesis, zac2022schemathesis}: on average, 52\% (99\%)
gains in statement (branch) coverage for EvoMaster and 29\% (75\%) statement
(branch) coverage gains for Schemathesis.
Finally, for OpenAPI specification inference, our results show that the
technique 
computes API endpoints (or resource paths) with 98\% precision and
56\% recall against the ground truth of existing API specifications.
These results demonstrate the benefits of our technique.

The contributions of this work are:

\begin{itemize}

  \item A first-of-its-kind approach for carving API test cases from UI paths that enables
    API-level testing for web applications, irrespective of the
    frameworks they use.

  \item A novel technique for inferring API specifications for web applications
    that use RESTful services.

  \item An implementation of the techniques in a tool called \toolname{} that is
    publicly available~\cite{toollink}. 

  \item Empirical assessment of \toolname, demonstrating the tool's
    effectiveness and the benefits of carved tests.



\end{itemize}

\section{Background and Motivating Example}
\label{sec:background}



\begin{figure}[t]
  \vspace*{-9pt}
  \begin{lstlisting}[language=Java,
%      basicstyle=\tiny,
      caption={Example OpenAPI specification.},
      label={listing:openAPISpec}
    ]
info:
  title: Conduit API
servers:f
  - url: http:/localhost:3000/api
paths:
/articles/{id}:          /* path item */
    get:                 /* HTTP method */
      parameters:
        - name: id
          in: path
          required: true
          schema: Integer
          example: 2
      responses:
        200:
          description: OK
          content:
            application/json:					/* MIME */
              schema:
                ref: '/schemas/SingleArticleResponse'
\end{lstlisting}
\end{figure}

\begin{figure*}[!t]
\centering
\includegraphics[trim=0.5cm 0cm 0.5cm 0cm, clip=true, width=0.9\linewidth
]{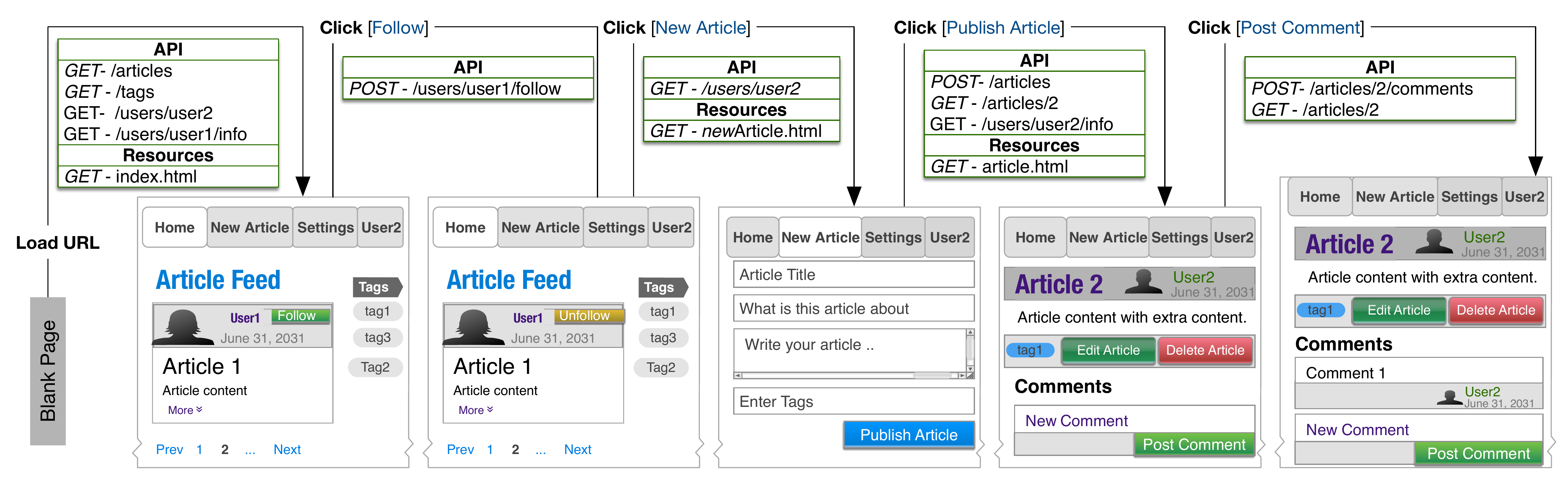}
\vspace*{-4pt}
\caption{Example illustrating a sequence of UI actions and states along
  with the API calls that are triggered by UI events.}
\vspace*{-14pt}
\label{fig:motivatingExample}
\end{figure*}

REST APIs~\cite{fielding2000architectural} are typically described in a specification (e.g., in OpenAPI~\cite{openapi} format, previously known as Swagger) that lists the available service operations, the input and output data
structures for each operation, and the possible response codes.
%
\autoref{listing:openAPISpec} shows a snippet of the OpenAPI spec for
the REST API of a web application called \texttt{\footnotesize
  realworld}~\cite{realworld} (one of the applications used in our
evaluation). The spec lists path items (under \texttt{\footnotesize paths:}),
where a \textit{path item} consists of a resource path (also referred to as
``API endpoint'') together with one or more HTTP methods~(or ``Operations''). The path item
illustrated in \autoref{listing:openAPISpec} shows the resource path (line~6),
the HTTP method (line~7), the parameters specification (lines~8--13), and the
response specification (lines 14--20). The response specification lists the
status code (line~15) and the response data format (line~18) and the structure
(lines~19--20). The structure definition contains a reference to a schema
defined elsewhere in the document (omitted here).

The resource path \texttt{\footnotesize /articles/\{id\}} (line~6) is specified as a
URI template~\cite{uriTemplate}, with path parameter \texttt{\footnotesize id}. Such a
resource path describes a range of concrete URIs via parameter expansion. A
concrete URI instance in an HTTP request targeting that endpoint contains an
integer value for \texttt{\footnotesize{id}} (e.g., \texttt{\footnotesize
  /articles/2}).  More generally, a path item can contain four kinds of
parameters---path, query, cookie, and header. Path and query parameters are
related to the URI, whereas header and cookie parameters are associated with
HTTP request headers.

\label{sec:example}

 \autoref{fig:motivatingExample} shows a UI test path for the
 \texttt{\footnotesize realworld}~\cite{realworld} web application. The test
 performs five UI actions that navigate through different application
 states. Each action exercises a specific functionality. For example,
 ``\texttt{\footnotesize {\color{gray}Click}[{\color{blue}follow}]}'' invokes
 the functionality to \emph{follow} a given user. The figure also shows the
 server-side APIs invoked by the browser for each UI action. The UI
 states are then updated based on the server response.
 For
 instance, ``\texttt{\footnotesize {\color{gray}Click}[{\color{blue}follow}]}''
 invokes the API ``\texttt{\footnotesize
   {\color{gray}POST[{\color{darkgray}/users/user1/follow}]}}'' and the UI state
 is updated, to show that ``follow user'' succeeded.

From the perspective of functional testing of the server-side APIs of a web
application, the UI actions and the API calls invoke the same functionality and,
therefore, would have the same code coverage and fault-detection
abilities. However, invoking the APIs directly, instead of going through the UI
layer has advantages: API calls exercise the service-side functionality much
more efficiently and are less prone to the brittleness usually associated with
UI tests~\cite{Grechanik:2009:MEG:1555001.1555055}, while exercising the APIs in
the manner they are invoked from the UI. Thus, carved API tests can be
convenient for developers to use in the course of their development
activities. This is not to say that carved API tests are an alternative to, or
replacement for, the UI tests. UI testing has an important role to play in
covering end-to-end flows through all the application tiers; however, such
testing is more suitable for system-level or acceptance testing in practice, and
less so for supporting developers in their server-side development activities.
Our first goal in this work, therefore, is to enable API-level testing such that
it is universally applicable for all web applications irrespective of the web
frameworks they use.

The second goal of our work is to infer an API specification, such as the one
illustrated in \autoref{listing:openAPISpec},
automatically for the server-side APIs of a web application. The inferred
specification documents the APIs and can also be used as input to automated API
testing tools (e.g.,~\cite{arcuri2019restful, atlidakis2019restler,
  viglianisi2020resttestgen, martin2020restest,Zac2021schemathesis}). 
API specification inference is
applicable to web applications that implement RESTful APIs.  Although API
specifications could also be inferred for other types of web applications and
could serve as useful documentation of server-side APIs, they would be less
effective as inputs to automated API testing tools.

The core challenge in specification inference is how to compute resource paths
with path parameters accurately (e.g., the \texttt{\footnotesize \{id\}} component of
resource path \texttt{\footnotesize /articles/\{id\}}). The concrete URI instances in
the requests observed at runtime contain integer values for \texttt{\footnotesize id},
such as \texttt{\footnotesize /articles/2}. The technique has to determine
which segments of concrete URIs represent path parameters. Moreover, a resource
path can have multiple path parameters, which adds to the complexity of the
problem.  In the next section, we present a dynamic-analysis-based carving
technique for addressing these challenges.

\section{Approach}
\label{sec:approach}

\begin{figure}[!t]
\centering
\includegraphics[trim=0cm 0cm 0cm 0cm, clip=true, width=0.85\columnwidth
]{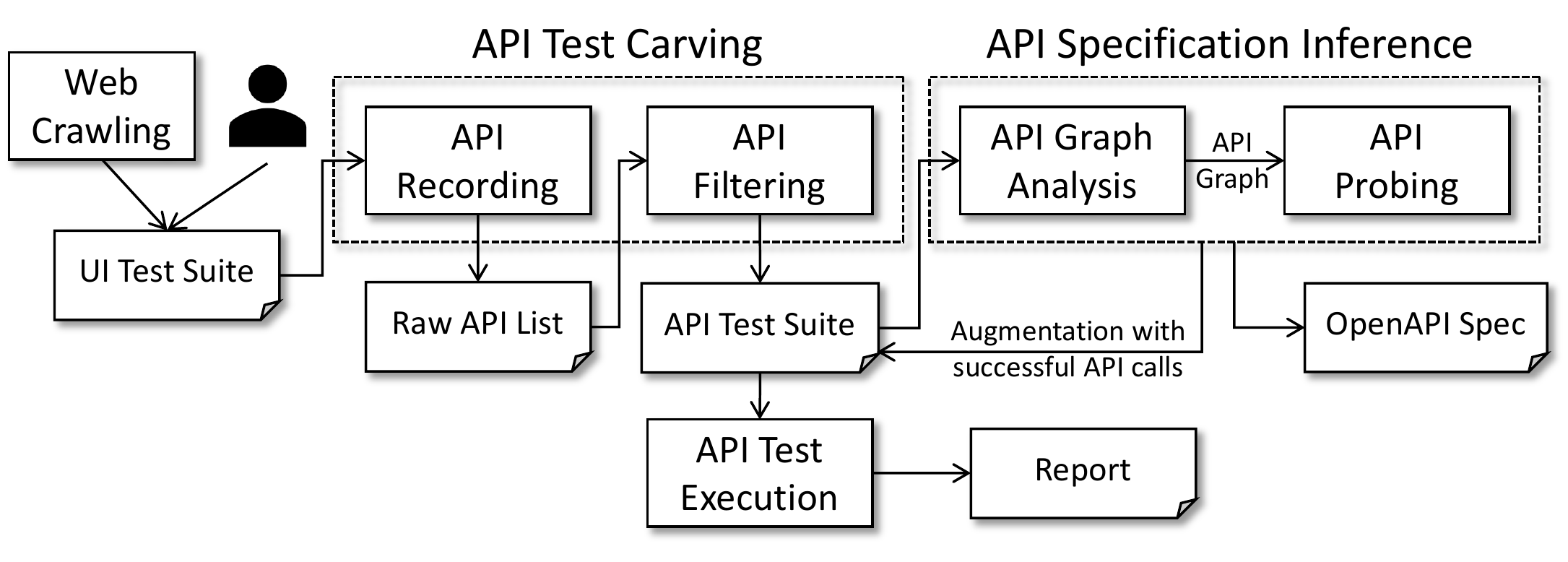}
\vspace*{-7pt}
\caption{Overview of our technique \toolname.}
\label{fig:technique}
\end{figure}

\autoref{fig:technique} presents an overview of our technique called \toolname. The input is a suite of UI test cases for a web application---the test cases could be automatically generated (e.g., created via automated web crawling) or
implemented by developers. The output consists of an
API-level test suite, along with a test-execution report, and for applications that use RESTful APIs, an OpenAPI
specification describing the server-side APIs of the web application.  The API
test suite is composed of carved test cases that are augmented with API calls
made during specification inference. The test-carving phase of the technique
involves API recording and API filtering. The specification-inference phase
constructs an API graph from the API test suite, and analyzes the graph to
create an OpenAPI specification. A key step during specification inference is
\emph{API probing}, which attempts to expand the set of resource paths observed during
UI test execution and discover additional information for creating more
accurate specifications as well as augmenting the carved test suite.  Next, we
describe the two phases of the technique in detail.

\subsection{API Test Carving}
\label{sec:test-carving}

\toolname performs API test carving in two steps.  In the first step, API
recording, the technique monitors API calls that are triggered through the execution of the UI test suite and logs the raw API calls. 
To record API calls, we add network listeners to the
browser executing the UI tests, which capture the raw outgoing and incoming HTTP traffic. As
~\autoref{fig:motivatingExample} illustrates, a UI action can result in multiple
API calls being executed, e.g., \texttt{\footnotesize
  {\color{gray}Click}[{\color{blue}Publish Article}]} triggers one \textsc{post}  and three
\textsc{get} requests. These requests,
together with their corresponding responses, are logged during API recording.

In the second step, API filtering, the technique applies a series of
filters---operation filter, status filter, and MIME filter---to the raw API
calls to remove the calls that are irrelevant for API test and specification
carving. The \textit{operation filter} is based on HTTP method checking and is
designed to omit methods that are unrelated to resource manipulation. This
filter removes all calls with HTTP methods \textsc{trace} and
\textsc{connect}. The \textit{status filter} checks the response status codes and excludes calls with unsuccessful requests, indicated by 4xx and
5xx response codes. Finally, the \textit{MIME filter} checks the MIME type of
the response payload and retains only those calls whose response payloads
contain JSON or XML data (i.e., MIME types \texttt{\footnotesize text/json} or
\texttt{\footnotesize text/xml}). For example, the resource-related API calls shown in~\autoref{fig:motivatingExample}, which are irrelevant for API-level
testing, are removed during filtering.



In the implementation of our technique, the filtering step is configurable. The
three filters described here proved to be adequate for our
experimentation. However, the user can configure filtering to prevent omission
of certain requests considered essential for API testing or provide custom
filters to omit additional types of API calls not covered by the three filters.
Filtering configuration may also be needed based on web application
characteristics; e.g., the MIME filter would be relevant if the application
consists of RESTful APIs.

The output of the filtering step consists of sequences of API calls from which
the carved API test suite is created. 



\subsection{API Specification Inference}
\label{sec:spec-inference}

\newcommand{\inferspec}{\texttt{\footnotesize InferSpec}\xspace}

\begin{figure}[!t]
\centering
\includegraphics[trim=0cm 0cm 0cm 0cm, clip=true, width=0.8\columnwidth
]{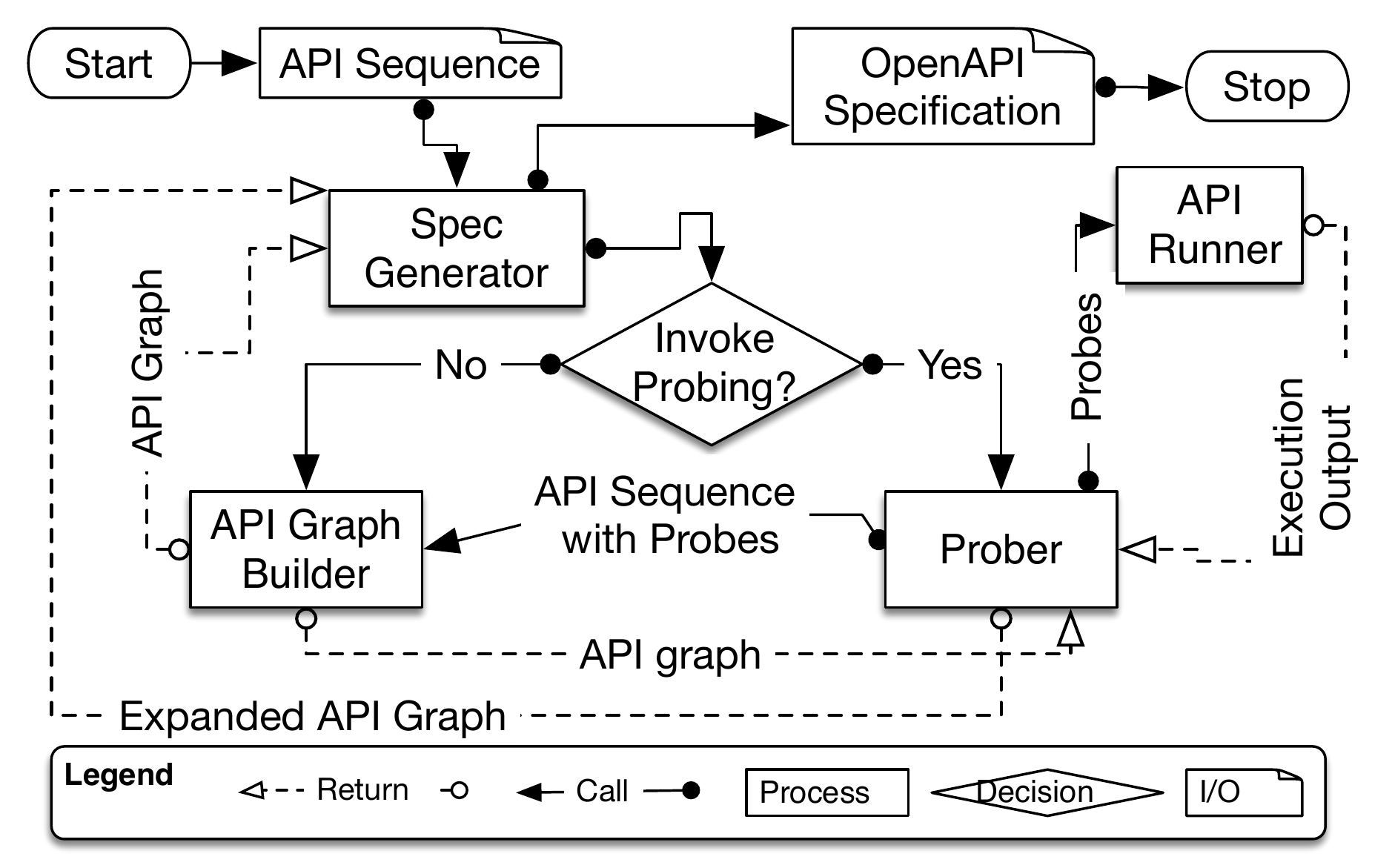}
\vspace*{-5pt}
\caption{The specification-inference flow (\inferspec).}
\label{fig:SpecInference}
\end{figure}

As discussed in \ref{sec:background}, the core problem in specification
inference that our approach addresses is computing path parameters for resource
paths. The technique has to detect the path components of concrete URI instances
that represent parameters, while handling paths with multiple parameters and
compensating for server-side state changes as a result of UI actions that can potentially impact server responses for URIs.

\autoref{fig:SpecInference} presents a flow chart, \inferspec, illustrating
specification inference. \inferspec takes as input the API sequences created by
API carving and produces as output an API specification. It builds an API graph
to represent the discovered resource paths and analyzes the graph to create the
API specification. \inferspec also analyzes the graph to generate
API probes (i.e., concrete API requests) that are executed against the
application to discover additional valid API calls that are missing in the
initial set of API sequences.\footnote{In the implementation of the technique, API probing
can be limited by upper bound on exploration time or number of probes executed.}
This step is intended to address the incompleteness of the initial API sequences
and, thereby, improve the accuracy of the inferred API specification. As an
additional benefit, the successful probes can be used for augmenting the carved API
test suite, potentially increasing its coverage.


\begin{algorithm}[t]
\scriptsize
\SetAlgoLined
\DontPrintSemicolon
\SetNoFillComment
\SetKwProg{Fn}{Function}{:}{end}
   \Fn{BuildAPIGraph}{
   \SetKwInput{Kw}{Init}
   \KwIn{apiset $\gets$ [\apiCall{1}...\apiCall{n}]\tcc*{Set of API Calls }}
   \Kw{\apiGraph{} $\gets$ $\phi$\tcc*[f]{API Graph for the given set of API calls}}
   \ForEach(){\apiCall{} $\in$ apiset }{
   	$path$ $\gets$ \apiCall{}.\request{}.URL.path\;
	SegArray $\gets$ $path$.split() \tcc*[f]{split path into segments}\;
	parent $\gets$ $root$         \tcc*[f]{a dummy starting node}\;
	 \ForEach(){Seg$_i$ $\in$ SegArray}{
	 	parentPath $\gets$ join(Seg$_0, \ldots,$ Seg$_{i-1}$)  \;
		{$v \gets \phi$ \tcc*{path parameter inference later}}
		\uIf{$i$ = SegArray.size}{
			end-point $\gets$ True \;
		}
		\Else{
			end-point $\gets$ False \;
		}
	 	{pathSeg $\gets$ (Seg$_i$, $i$,  parentPath, SegArray.size, endpoint, $v$)}\;
		segExists $\gets$ for-all \pathNode{s} $\in$ $\mathcal{N}$ AreEqual(pathSeg, \pathNode{s})\;
		\uIf {segExists}{
			\apiGraph{}.addEdge(parentNode, \pathNode{sim}) \;
			parentNode = \pathNode{sim} \;
		}
		\Else{
			\apiGraph{}.addNode(pathSeg) \;
			\apiGraph{}.addEdge(parentNode, pathSeg) \;
			parentNode = pathSeg \;
		}
	 }
   }
   \KwRet{\apiGraph{}}
}
\Fn{AreEqual~({\pathNode{1}, \pathNode{2}{})}}{
  \KwOut{True or False} 
		\If{(\pathNode{1}$.n$ $\neq$ \pathNode{2}$.n$) $\lor$ (\pathNode{1}$.d$ $\neq$ \pathNode{2}$.d$) } 
		{
		    \textbf{return} False  \tcc*[f]{different name or path index}\;
		}
		
	    \If{\pathNode{1}$.p$ $=$ \pathNode{2}$.p$}{
	        \textbf{return} True  \tcc*[f]{same name, path index, and parent path}\;
	    }
	    
	    \eIf{IsEndPoint(\pathNode{1}) $\wedge$ IsEndPoint(\pathNode{2})}
	    {
	        \textbf{return} CompareResponses(\pathNode{1}$.l$,
                \pathNode{2}$.l$)\;
	    }
	    {
	        \textbf{return} False  \tcc*[f]{different parent path}\;
	    }
		
}
\caption{API graph construction}
\label{algo:apiGraph}
\end{algorithm}

\subsubsection{API Graph Construction}
 
Algorithm~\ref{algo:apiGraph} presents the algorithm for building the API graph.
Before describing the algorithm, we introduce some terminology.


\begin{small}
\begin{definition}[\textbf{Path Segment}]\label{def:def_pathNode}
Given a URI $U$, a \textit{path segment} \pathNode{} is a tuple~($n$, $d$, $p$,
$e$, $l$, $v$), where $n$ is the segment string, $d$ is the index of the segment
in $U$, $p$ is the parent path for \pathNode{}, $e$ is a boolean indicating
whether \pathNode{} is the final segment of $U$ (and, therefore, an API
endpoint), $l$ is the response payload (if $e$ is true), and $v$ is the path
parameter inference result for \pathNode{}.
\end{definition}

\begin{definition}[\textbf{API Graph}]\label{def:def_apiGraph}
An \textit{API graph} \apiGraph{} $=$ (\pathNode{r}, $\mathcal{N}$,
$\mathcal{E}$) is a directed acyclic graph, where \pathNode{r} is the dummy root
node of the graph, $\mathcal{N}$ is a set of nodes, and $\mathcal{E}$ is a set
of edges.  Each node in $\mathcal{N}$ is a path segment and a pair of
consecutive segments in a URI is connected by an edge in $\mathcal{E}$.
\end{definition}
 
\begin{definition}[\textbf{Graph Path}]\label{def:def_graphPath}
A \textit{graph path} in an API Graph is a sequence of path
segments~(\pathNode{r}$,\ldots,$ \pathNode{x}) that connects the root
node~\pathNode{r} to any graph node \pathNode{x}. A complete path is a path from
\pathNode{r} to a node where \pathNode{}.$e$ is true.
\end{definition}
\end{small}

\begin{figure}[t]
\begin{subfigure}[b]{\columnwidth}
\includegraphics[trim=0cm 0cm 0cm 0cm, clip=true,
  width=\textwidth]{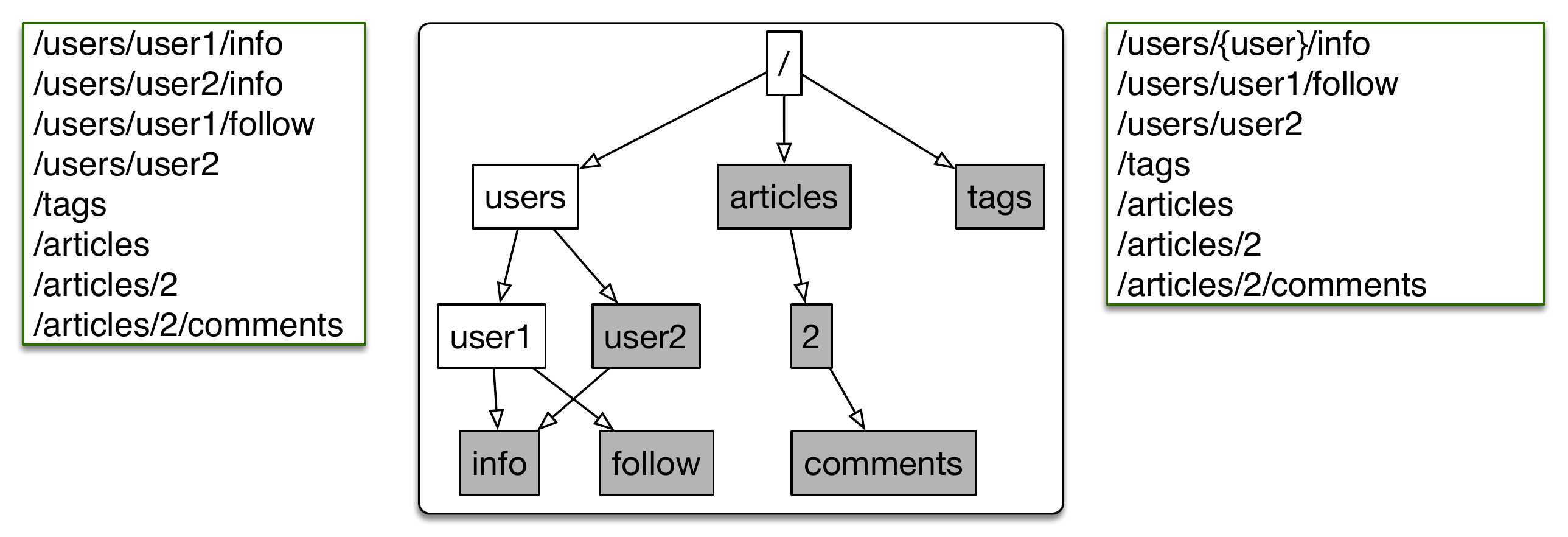}
\vskip -6pt
\caption{\footnotesize API calls carved from the UI execution}
\label{fig:fromUI}
\end{subfigure}

\begin{subfigure}[b]{\columnwidth}
\includegraphics[trim=0cm 0cm 0cm 0cm, clip=true, width=\textwidth]{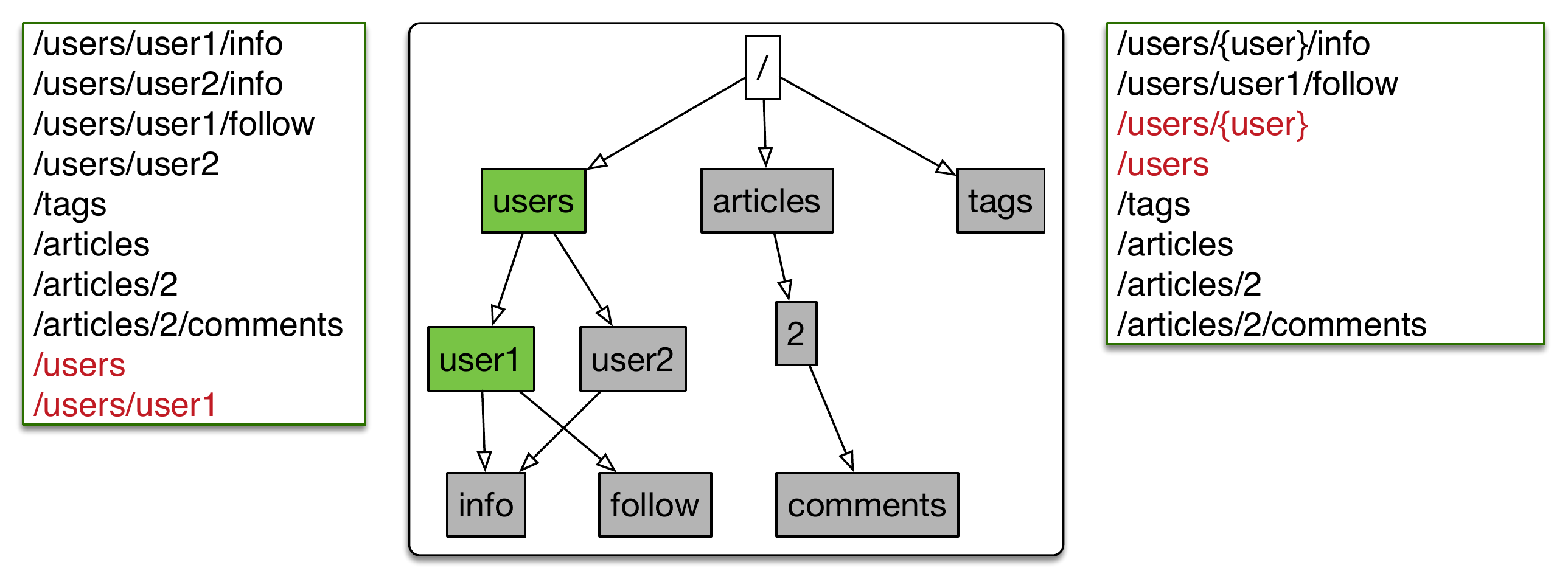}
\vskip -6pt
\caption{\footnotesize Probing Stage 1: probes for intermediate nodes}
\label{fig:intermediate}
\end{subfigure}

\begin{subfigure}[b]{\columnwidth}
\includegraphics[trim=0cm 0cm 0cm 0cm, clip=true, width=\textwidth]{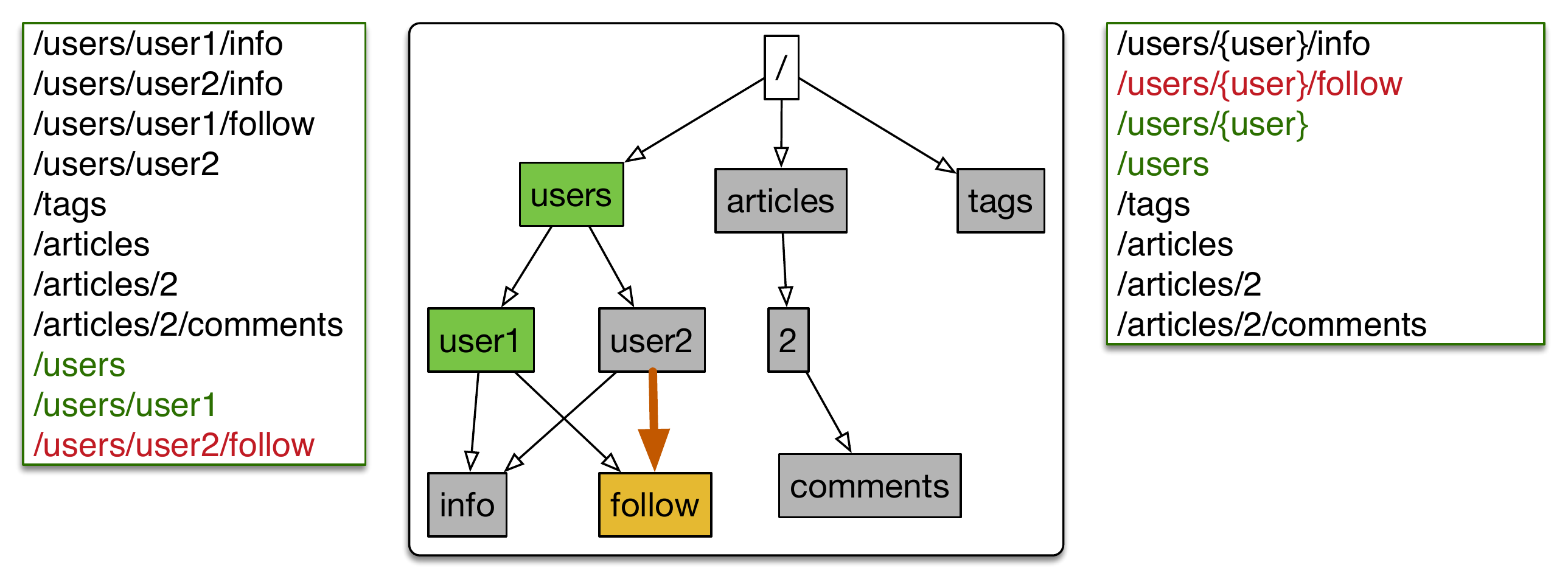}
\vskip -6pt
\caption{\footnotesize Probing Stage 2: probes from bipartite analysis}
\label{fig:bipartite}
\end{subfigure}

\begin{subfigure}[b]{\columnwidth}
\includegraphics[trim=0cm 0cm 0cm 0cm, clip=true, width=\columnwidth]{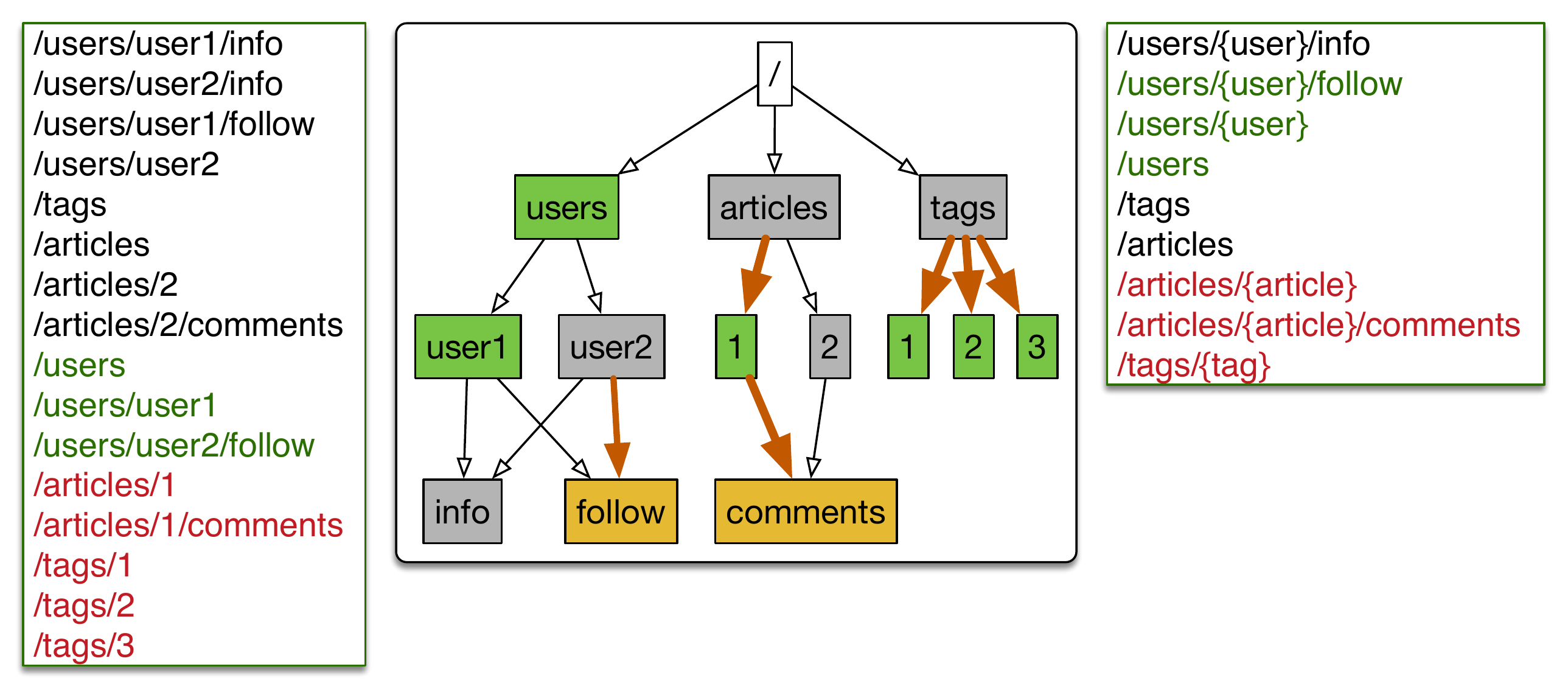}
\vskip -6pt
\caption{\footnotesize Probing Stage 3: probes from response analysis}
\label{fig:responseAnalysis}
\end{subfigure}

\footnotesize Each subfigure shows the API graph~(middle) built from API
calls~(left) and the inferred specification~(right). The color coding of API
calls (spec paths) indicates calls recorded (paths created) during UI navigation
(black) and probes (paths) created in the previous stage (green) and the current
stage (red). The color coding of nodes illustrates nodes created after UI
navigation (gray), nodes with responses discovered via probing (green), and
nodes with extra responses because of probing (yellow). The edge colors
highlight edges created by probes in the previous stage (green) and the current
stage (red).
  
%
\vspace*{-2pt}
\caption{Illustration of specification inference.}


\label{fig:techniqueExample}
\end{figure}

An API graph is constructed for a set of URIs (extracted from API calls). To illustrate, consider the API graphs shown in Figure~\ref{fig:techniqueExample}. Each API graph represents the URIs shown to the left of the graph. A graph node, except the root node, represents one or more segments from the URIs, and each complete path represents a URI.

Function \texttt{\footnotesize BuildAPIGraph} (lines~1--27) of
Algorithm~\ref{algo:apiGraph} iterates over a given set of API
calls~(\apiCall{}) and builds an API graph by parsing each request URI into a
path in the graph. For each URI, the algorithm splits the URI into segments and
then builds a path segment for each segment~(lines~7--14). If a path segment
\pathNode{} is not similar to any of the existing path segments in the graph,
\pathNode{} is added to the graph (lines~15--23).

The similarity of two path segments is determined by the function
\texttt{\footnotesize AreEqual} (lines~28--40), which first compares the names
and path indexes of the two segments (line~29).  If either of these do not
match, the path segments are considered to be different.  For example, as shown
in~\autoref{fig:techniqueExample}, each string segment in a URI has its own node
in the graph.  If the names and path indexes match, the function next compares
the parent paths of the segments (using string comparison) and considers the
segments to be equivalent if the parent paths match (lines~32--34). Otherwise,
if the segments represent endpoints, the function \texttt{\footnotesize
  CompareResponses} is called to determine segment equivalence (lines~35--36).
Note that a response object is available for a path segment only if there exists
an API call that ends at the segment, making it an endpoint.

\texttt{\footnotesize CompareResponses} (not shown in Algorithm~\ref{algo:apiGraph})
relies on the structural similarity of responses instead of matching the entire
responses. For a response with JSON or
XML data, it ignores the \emph{values} and builds a tree with \emph{keys} in
the data.  It then asserts the structural similarity of the trees to determine
 response similarity. For example, consider the requests [\texttt{\footnotesize
     GET /users/user1/info}] and [\texttt{\footnotesize GET /users/user2/info}]
 in~\autoref{fig:techniqueExample}) with responses \texttt{\footnotesize \{"id":1, "name":"user1",
   "role":"user"\}} and \texttt{\footnotesize \{"id":2, "name":"user2",
   "role":"user"\}}. To compare these two responses, the technique builds trees using the keys \texttt{\footnotesize [id,
     name, role]} and invokes a tree-comparison technique (APTED~\cite{apted}) to check their equivalence.

%

\begin{algorithm}[t]
\scriptsize
\SetAlgoLined
\DontPrintSemicolon
\SetNoFillComment
\SetKwProg{Fn}{Function}{:}{end}

   \Fn{ExtractOpenAPI}{
   \KwIn{\apiGraph{}\tcc*[f]{API Graph after path variable inference}}
   \KwOut{uriTemplates $\gets$ $\phi$}
   \apiGraph{} $\gets$ MergeLeafNodes(\apiGraph{})\;
   \ForEach(\pathNode{} $\in$ \apiGraph{}){ }{
   	\uIf{\pathNode{i}.$e$ = True}{
		paths $\gets$ GetGraphPaths(\pathNode{i}) \;
		\uIf{paths.size > 1}{
			template $\gets$ getURITemplate(paths)\;
		}
		\Else{
			template $\gets$ paths[0]
		}
		uriTemplates.add(template) \;
	}		
    }
   \KwRet{uriTemplates}
}
   \Fn{MergeLeafNodes}{
   \KwIn{\apiGraph{}\tcc*[f]{API Graph for the given set of API calls}}
   \ForEach(){ \pathNode{i} $\in$ \{\pathNode{n}\} }{
   	 \ForEach(){ \pathNode{j} $\in$ \{\pathNode{n}\} }{
	 \tcc{Assign a variable when nodes have matching responses}
	 	   \uIf{(\pathNode{i}.$e$=True) $\wedge$  (\pathNode{j}.$e$=True) $\wedge$   	(CompareResponses(\pathNode{i}, \pathNode{j})=True)}{
			\pathNode{i}.$v$ = \pathNode{j}.$v$ = variableMap.get() \;
		}
	 }
   }
   \KwRet{\apiGraph{}}
   
}
 \Fn{GetGraphPaths}{
  \KwIn{\apiGraph{}, \pathNode{x} \tcc*[f]{An API Graph and a node in it}}
  \KwOut{paths $\gets \phi$}
	  \ForEach(){ \graphPath{i} $\in$ \apiGraph{}.getPathsto(\pathNode{x})}{
	  	path $\gets$ $\phi$ \;
   		 \ForEach(){ \pathNode{i} $\in$ \{\pathNode{n}\} }{
			\uIf{\pathNode{i}.$v$ != $\phi$}{
				path.add(\pathNode{i}.name)\;
			} 
			\Else{
				path.add(\pathNode{i}.$v$) \tcc*[f]{$v$ is set  by MergeLeafNodes}\; 
			}
	 	}
		paths.add(path)\;
	}
	\KwRet{paths}\;
 }

%
\caption{Generating API specification from API graph}
\label{algo:specGen}
\end{algorithm}

\subsubsection{API Specification Generation}

Algorithm~\ref{algo:specGen} presents the steps involved in generating API
specification from the API graph.
Our goal in specification inference is to make the specification precise in
terms of the number of path items for each API endpoint.
An ideal specification
should have exactly one path item describing an API endpoint; URI templates
with path parameters make it possible to do so.

The algorithm first merges leaf nodes in the API graph (line~2). The function \texttt{\footnotesize MergeLeafNodes} (lines~15--23) performs response comparison to
determine whether two nodes with different names belong to the same API
endpoint. In addition, the algorithm makes use of the graph structure by getting paths that
reach the same endpoint node in the API graph (lines~24--37).  Finally, after
computing a list of URIs that belong to similar endpoint nodes in the API graph,
the technique performs a simple path-index-based match per URI segment to extract a
template (lines~5--11).

For example, in~\autoref{fig:fromUI}, for the leaf node \code{info}, \code{GetGraphPaths} returns two  graph paths for URIs
\texttt{\footnotesize /users/user1/info} and \texttt{\footnotesize
  /users/user2/info}, which the technique considers equivalent and extracts the template   
\texttt{\footnotesize
  /users/\{user\}/info} with path parameter \texttt{\footnotesize \{user\}}.

For the example in~\autoref{fig:intermediate}, the graph paths for URIs
\texttt{\footnotesize /users/user1} and \texttt{\footnotesize /users/user2} end
at different segments. However, \texttt{\footnotesize MergeLeafNodes} performs
response comparison to determine that these segments are equivalent and sets
a variable to represent the path parameter (line~19 of Algorithm~\ref{algo:specGen}). Then, \texttt{\footnotesize GetGraphPaths} uses that
variable and returns the string \texttt{\footnotesize /users/\{user\}} for both
URIs. Finally, that returned string is used as the template with path parameter
\texttt{\footnotesize \{user\}}. We use variable name \texttt{\footnotesize
  user} here for readability; our implementation creates variable names such as
\texttt{\footnotesize var0}.


\subsubsection{API Graph Expansion via Probing}

The API graph created from the set of API calls seen during UI navigation is
limited by the completeness of UI tests, 
which can affect the precision and completeness of the inferred
API specification.  To address this, \toolname expands the initial API graph
via systematic API probing.


The technique creates four types of probes: intermediate,
bipartite, response, and operation. Intermediate and bipartite probes are
built via API graph analysis, response probes are based on HTTP response
analysis, and operation probes aim to discover unseen operations for known API
endpoints. After building probes, the technique sends them to the server using a
scheduling algorithm that avoids data dependencies. The successful probes (i.e.,
probes with response codes other than 4xx or 5xx) are used for enhancing the API
graph and augmenting the API test suite. \inferspec uses the expanded API graph
to generate a potentially more accurate API specification.


\noindent
\paragraph*{Intermediate probes}
These probes are created for API graph nodes that do not have an associated
server response (i.e. \pathNode{}$.e$ is false). For example,
in~\autoref{fig:intermediate}, the probes \texttt{\footnotesize /users}
and \texttt{\footnotesize /users/user1} are intermediate probes 
built for the nodes \texttt{\footnotesize users} and \texttt{\footnotesize user1} that do not have an associated response. 
In this case, the two
endpoints are indeed valid. As a result, \inferspec adds a new path
item \texttt{\footnotesize /users} and computes path
variable \texttt{\footnotesize user} for the existing path
\texttt{\footnotesize /users/user2}, which it replaces with
the template \texttt{\footnotesize /users/\{user\}}.

\begin{figure}[!t]
\vspace*{-4pt}
\centering
\includegraphics[trim=0cm 0cm 0cm 0cm, clip=true, width=0.85\linewidth
]{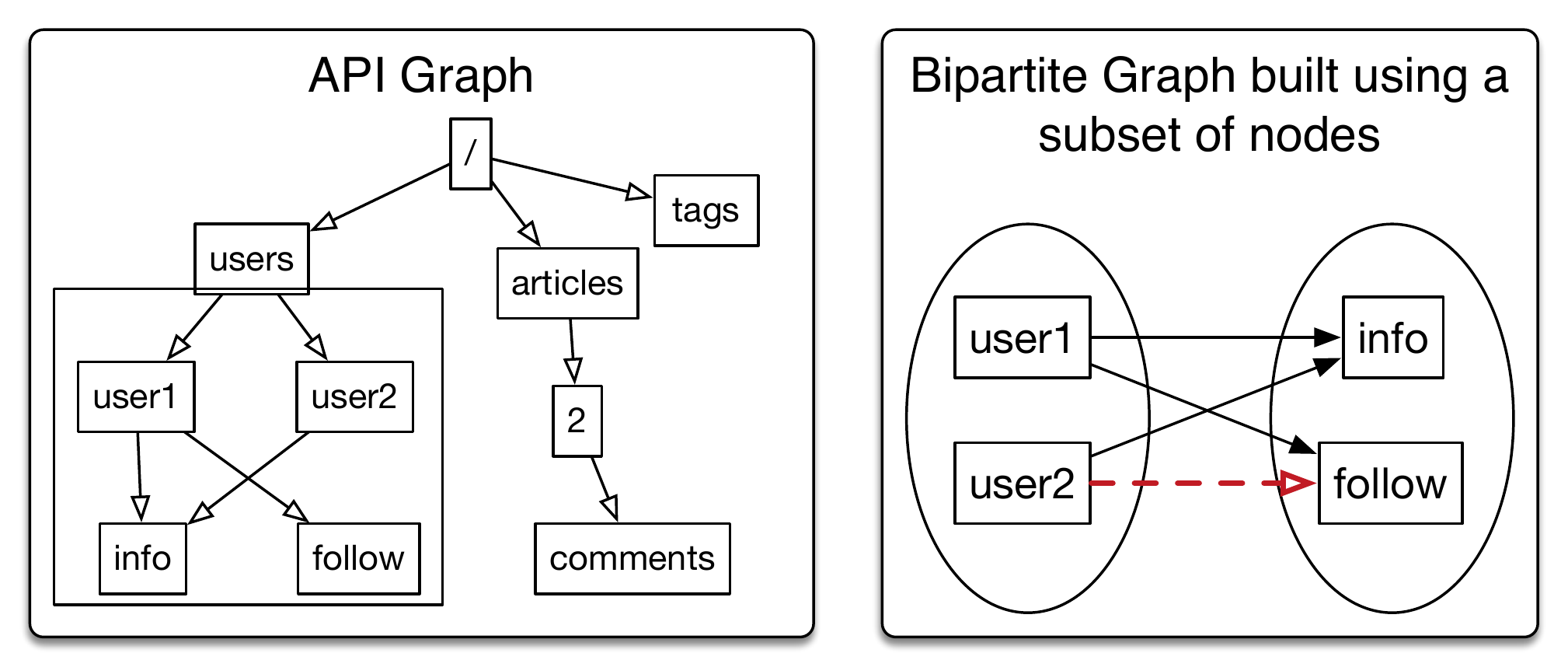}
\vspace*{-4pt}
\caption{Bipartite analysis for API probe generation.}
\label{fig:bipartiteGraph}
\end{figure}

\paragraph*{Bipartite probes}
These probes are generated by building a bipartite graph from join nodes (i.e.,
nodes that have more than one predecessor) in the API graph. To illustrate,
consider the example in~\autoref{fig:bipartiteGraph}, where join
node \texttt{\footnotesize info} has two predecessors (\texttt{\footnotesize
user1} and \texttt{\footnotesize user2}). For this node, the technique
constructs the bipartite graph shown in the figure: the left part of the graph
contains all predecessors of the join node and the right part contains all
successors of nodes in the left part. The technique then computes missing edges
that would make the bipartite graph complete, i.e., each node on the left is
connected to each node on the right. In this example, one missing edge makes the
bipartite graph complete. From this analysis, probe \texttt{\footnotesize
/users/user2/follow} is generated. As shown in~\autoref{fig:bipartite}, this new
probe lets the technique infer the new path variable \texttt{\footnotesize user} and
convert concrete resource path \texttt{\footnotesize /users/user1/follow} to
path template \texttt{\footnotesize /users/\{user\}/follow}, thereby improving
the specification.


\paragraph*{Response probes}
Response probes are generated by analyzing the server responses for existing API
calls. For each response object, the technique builds probes from keys and
values extracted from the response. Suppose the response
for \texttt{\footnotesize GET /tags} is \texttt{\footnotesize [\{ id: 1, name:
tag1, author: user1\}, \{ id: 2, name: tag2, author: user1\} ..]}. Using this response, we build probes such
as \texttt{\footnotesize /tags/1, /tags/id, /tags/tag2, /tags/author}.
As shown in~\autoref{fig:responseAnalysis}, by analyzing
the \texttt{\footnotesize articles} object, we build
probes \texttt{\footnotesize /articles/1/comments} and \texttt{\footnotesize
/articles/1}, which result in inference of path templates \texttt{\footnotesize
/articles/\{article\}} and \texttt{\footnotesize
/articles/\{article\}/comments}. Similarly, response analysis
of \texttt{\footnotesize tags} object helps us infer \texttt{\footnotesize /tags/\{tag\}}.

\paragraph*{Operation probes}
Operation probes are generated by analyzing API calls based on coverage of HTTP
methods per known API endpoint. We consider seven HTTP
operations---\textsc{get}, \textsc{post}, \textsc{put}, \textsc{patch}, \textsc{options}, \textsc{head},
and \textsc{delete}---in our analysis. For example, if the existing set of API
calls contains \texttt{\footnotesize GET /tags/1} and \texttt{\footnotesize
PATCH /tags/1}, we generate five probes for the endpoint, each covering one of the remaining HTTP operations (e.g., \texttt{\footnotesize DELETE /tags/1}).

\paragraph*{Probe Scheduling}
HTTP requests can cause server-side state updates and, in general, the server
response for a request can vary based on other requests.  Moreover, the
resources corresponding to a URI could be dynamic and only available in certain
server-side states. Therefore, API probing should be performed at appropriate
server states; our technique achieves this via probe scheduling, based on API graph
analysis.

\begin{figure}[t]
\centering
\includegraphics[trim=0cm 0cm 0cm 0cm, clip=true, width=0.85\linewidth
]{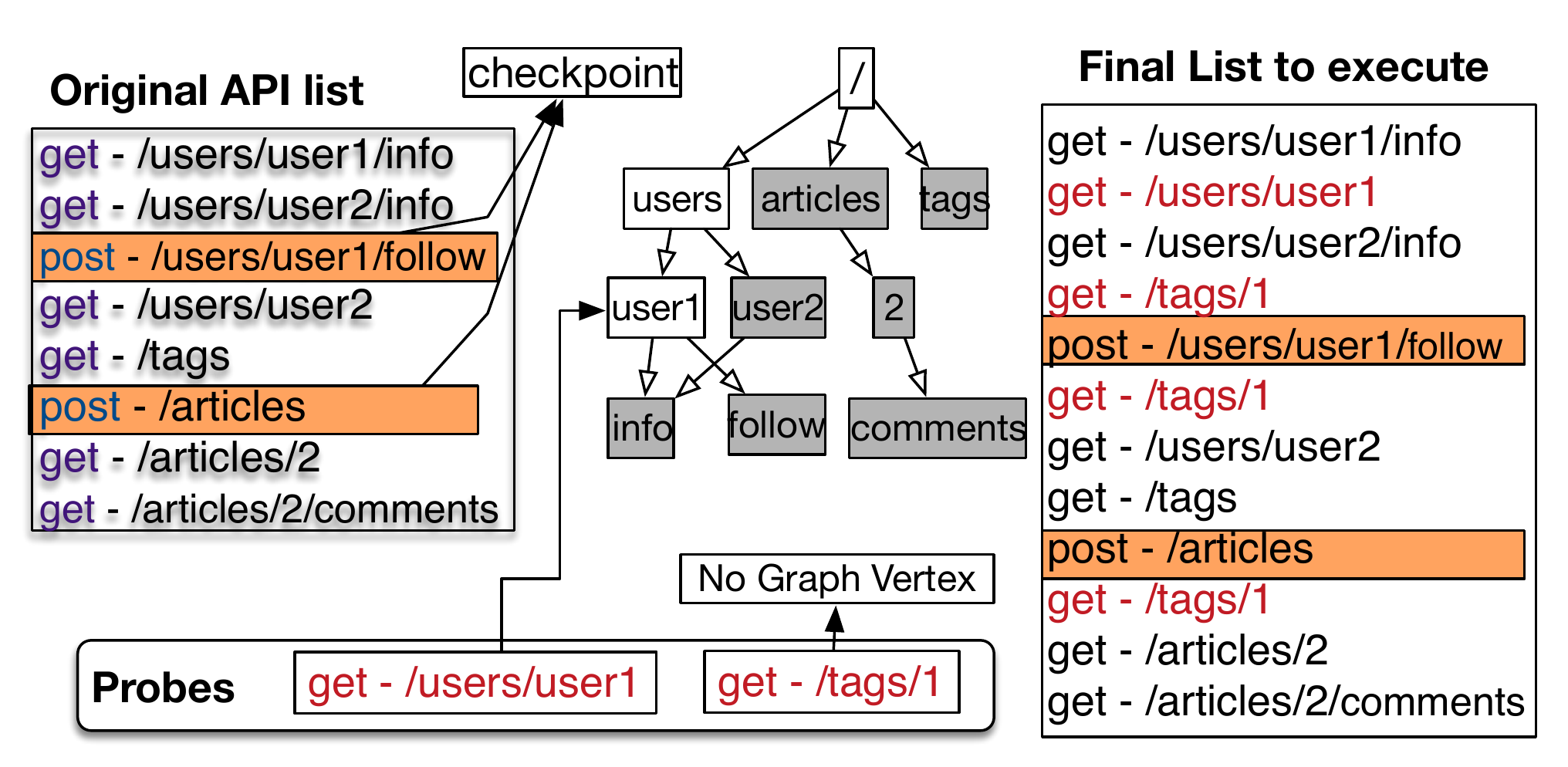}
\vspace*{-4pt}
\caption{Example for illustrating probe scheduling.}
\label{fig:probeScheduling}
\end{figure}

For each probe, we first check if the URI has a corresponding endpoint node in
the graph. Consider the example in~\autoref{fig:probeScheduling}.  Node
\texttt{\footnotesize user1}, which is the endpoint node for URI
\texttt{\footnotesize /users/user1}, already exists in the graph, whereas the
corresponding node for \texttt{\footnotesize /tags/1} does not exist.  If the
endpoint node for a probe exists, we schedule the probe immediately before the
last request in the original list whose URI includes the segment/node.  If the
endpoint node does not exist, we schedule the probe after each checkpoint
request. A \textit{checkpoint request} is an HTTP request that can change the
server-side state. We define two types of checkpoints: cookie-based and
operation-based. Cookie-based checkpoints are the HTTP requests for which the
server responds with a \code{set-cookie} field. Operation-based checkpoints correspond
to HTTP requests capable of modifying resources, i.e., HTTP methods
\textsc{put}, \textsc{post}, \textsc{delete}, and
\textsc{patch}. In~\autoref{fig:probeScheduling}, the two \textsc{POST} requests
are checkpoint requests. As the final list in~\autoref{fig:probeScheduling}
shows, \texttt{\footnotesize GET /users/user1} is scheduled only once,
immediately after the node is discovered in the graph, whereas
\texttt{\footnotesize GET /tags/1} is scheduled three times (corresponding to
three potential server-side states), once before any checkpoint and once each
after the two checkpoints in the original list.
After the probes are executed, we keep
only one instance of a successful probe in cases where multiple instances
succeed.

\section{Implementation}

We implemented our technique in a tool called \toolname.
We use 
Crawljax~\cite{crawljax} to generate~\cite{fraggen} UI test cases automatically. 
The API recorder module uses the Chrome Devtools
Protocol~\cite{cdp} along with Selenium~\cite{selenium} to instrument the
browser during UI test execution to record API calls. 
%
%
%
Our implementation and experimental dataset are publicly available in a replication
package~\cite{toollink}. 

\section{Empirical Evaluation}
\label{sec:evaluation}

We investigated the following research questions in the evaluation of \toolname.


%
%
%
%

\begin{description}

\item{\q{1}}: How do carved API tests compare with UI tests in terms of code coverage and execution efficiency?

\item{\q{2}}: How effective is \toolname in generating OpenAPI specifications?

\item{\q{3}}: Do carved API tests improve the coverage achieved by automatically generated API test suites?


\end{description}

\subsection{Experiment Setup}

We performed the evaluation on seven open-source web applications;
\autoref{table:table_subjects} lists the applications and their
characteristics. All of the applications implement RESTful APIs for their
services and have OpenAPI specifications available, which serve as ground truth
for measuring the accuracy of the inferred API specifications.

\begin{table}[]
\renewcommand{\arraystretch}{1.05}
\caption{\small Web applications used in the evaluation.}
\vspace*{-5pt}
\centering
\label{table:table_subjects}
\resizebox{\columnwidth}{!}{

\begin{tabular}{llrrr}
\toprule
Application & Framework              & \# API Endpoints & \# Operations & LOC  \\
\midrule
booker      & Spring-boot, ReactJS   & 15               & 24         & 8K   \\
ecomm       & Spring-boot            & 21               & 22         & 6K   \\
jawa        & Spring-boot, AngularJS & 5                & 8          & 20K  \\
medical     & Spring-boot, VueJS     & 20               & 28         & 5K   \\
parabank    & Spring-mvc, AngularJS  & 27               & 27         & 60K  \\
petclinic   & Spring-boot, AngularJS & 17               & 36         & 39K  \\
realworld   & Express, NextJS        & 12               & 19         & 12K  \\
\midrule
Total       &                        & 117              & 164        & 150K\\
\bottomrule
\end{tabular}
}
\end{table}

\begin{table*}[t]
\caption{Statistics about different analysis stages in \toolname runs on the subject applications.}
\vspace*{-5pt}  
\label{table:toolStats}
\footnotesize
\renewcommand{\arraystretch}{1.05}
\resizebox{0.99\linewidth}{!}{
\begin{tabular}{l@{\hskip0.5em}r@{\hskip0.5em}r@{\hskip1.5em}r@{\hskip0.5em}r@{\hskip0.5em}r@{\hskip0.5em}r@{\hskip1em}r@{\hskip0.5em}r@{\hskip0.5em}r@{\hskip0.5em}r@{\hskip0.5em}r@{\hskip0.5em}r@{\hskip0.5em}r@{\hskip0.5em}r}
\toprule
          & \multicolumn{2}{c}{\multirow{2}{*}{API Filtering}} & \multicolumn{3}{c}{\multirow{2}{*}{Probing}}                               & \multicolumn{8}{c}{Generated Test Suites}                                                                                                                                                                                                                   \\ \cmidrule(lr){8-15}
          & \multicolumn{2}{c}{}                               & \multicolumn{4}{c}{}                                                       & \multicolumn{4}{c}{Carver}                                                                                         & \multicolumn{4}{c}{Carver + Prober}                                                                                                             \\ \cmidrule(lr){2-3}\cmidrule(lr){4-7}\cmidrule(lr){8-11} \cmidrule(lr){12-15}
       & Recorded       & \multicolumn{1}{l}{Filtered}      & Generated & \multicolumn{1}{l}{Executed} & Succeeded& \multicolumn{1}{l}{Checkpoints} & Paths total & \multicolumn{1}{l}{Paths success} & \multicolumn{1}{r}{Requests} & \multicolumn{1}{l}{Time (s)} & \multicolumn{1}{l}{Paths total} & \multicolumn{1}{l}{Paths success} & \multicolumn{1}{r}{Requests} & \multicolumn{1}{l}{Time (s)} \\
\midrule
booker    & 3610                         & 613                          & 529                           & 85295                        & 57                            & 203                             & 14                                        & 10                                & 443                          & 21                                 & 24                              & 20                                & 500                          & 21                                 \\
ecomm     & 7838                         & 1187                         & 440                           & 4906                         & 4                             & 18                              & 61                                        & 60                                & 1175                         & 21                                 & 61                              & 60                                & 1179                         & 15                                 \\
jawa      & 1568                         & 198                          & 60                            & 279                          & 13                            & 7                               & 9                                         & 4                                 & 110                          & 6                                  & 18                              & 13                                & 123                          & 6                                  \\
medical   & 315                          & 122                          & 1277                          & 39309                        & 17                            & 32                              & 24                                        & 23                                & 117                          & 15                                 & 26                              & 25                                & 134                          & 17                                 \\
parabank  & 13072                        & 574                          & 694                           & 43833                        & 26                            & 68                              & 25                                        & 23                                & 572                          & 125                                & 29                              & 27                                & 598                          & 124                                \\
petclinic & 1536                         & 294                          & 1399                          & 5144                         & 102                           & 42                              & 20                                        & 20                                & 290                          & 5                                  & 50                              & 50                                & 392                          & 7                                  \\
realworld & 1471                         & 398                          & 7510                          & 72259                        & 225                           & 9                               & 62                                        & 36                                & 365                          & 79                                 & 116                             & 91                                & 590                          & 103                                \\
\midrule
Total     & 29410                        & 3386                         & 11909                         & 251025                       & 444                           & 379                             & 215                                       & 176                               & 3072                         & 273                                & 324                             & 286                               & 3516                         & 294                               \\
\bottomrule

\end{tabular}
}
\vspace*{-12pt}
\end{table*}

For UI test generation, we configured Crawljax to run for 30~minutes. 
We also created 14 manual tests for three subjects (\texttt{\footnotesize booker}, \texttt{\footnotesize medical}, \texttt{\footnotesize ecomm}), which required dedicated action sequences and input data. 
Thus, our evaluation uses
automatically generated and developer-written UI test cases.

For investigating \q{3}, we used two popular automated test generators for REST
APIs---EvoMaster~\cite{evomaster} and
Schemathesis~\cite{schemathesis}. EvoMaster can be used in white-box and
black-box modes; in the white-box mode, it is applicable to REST APIs
implemented in the Java language. For our study, we used EvoMaster in its
black-box mode so that it  can be applied to non-Java API implementations in
our subjects. A recent empirical study~\cite{kim:2022} showed these two tools to
be the top-performing tools, in terms of code coverage achieved, among the
black-box testing tools for REST APIs. We configured EvoMaster to run
for one hour; for Schemathesis, we used its default configuration settings.  We
ran each tool 10~times to account for randomness and report coverage data
averaged over the 10~runs.
To measure code coverage, we used JaCoCo~\cite{jacoco} for Java-based APIs and
Istanbul~\cite{istanbul} for JavaScript-based APIs. 

\subsection{Quantitative Analysis of \toolname Stages}

Before discussing our results on the research questions, we present empirical
data on different stages of \toolname and provide a quantitative analysis of
the stages. \autoref{table:toolStats} presents data about the filtering,
probing, and test-generation stages.

Columns~2--3 of the table show the number of API calls available after recording
and filtering, and highlight the importance of filtering: i.e., a large
proportion of the raw API calls recorded get filtered out. These calls basically
retrieve resources related to UI rendering in the browser and can be ignored for
testing the functionality of server-side APIs. On average, over 88\% of the raw
API calls belong to the category of irrelevant calls. The proportion of such
calls ranges from over 72\% (for \texttt{\footnotesize realworld}) to over 95\%
(for \texttt{\footnotesize parabank}). Thus, API filtering is an important
component of \toolname; moreover, as discussed in
Section~\ref{sec:test-carving}, the filtering component can be configured to be
more strict (removing more of the raw API calls) or less stringent (removing
fewer calls).

Columns~4--7 present information about the probing stage: probes generated,
probes executed, and checkpoints in the filtered API list. On average, the
number of probes generated is over three times the number of filtered API calls,
and the number of probes executed is 21~times the number of generated
probes. Recall from the discussion of probe scheduling in
Section~\ref{sec:spec-inference} that some probes are scheduled for multiple
executions based on occurrences of checkpoints in the initial API list. The
number of generated probes varies considerably, ranging from 0.3~times the
initial API calls (for \texttt{\footnotesize jawa}) to almost 19~times the
initial API calls (for \texttt{\footnotesize realworld}). A similar large
variation can also be seen in the number of probes executed. Finally, 444 probes were successful and are added to API test suite generated at the end of probing. 

Columns~8--15 of~\autoref{table:toolStats} present data about test generation,
broken down by tests created during carving and probing. 
It can be seen that the total number of successful paths, which are the number of valid resource paths discovered, increases from 176 in the Carver test suite to 286 in the Prober test suite. The Prober is, thus, able to discover 110  additional valid resource paths across the applications. These 110 path invocations come at the cost of only 21 seconds. In other words, the Prober test suite is able to successfully exercise 62.5\% more paths with only 7.6\% increase in test-execution time. 

\begin{figure}[t]
\centering

\begin{subfigure}[b]{0.54\columnwidth}
\includegraphics[trim=0.3cm 0.5cm 0.2cm 1.5cm, clip=true, width=\textwidth]{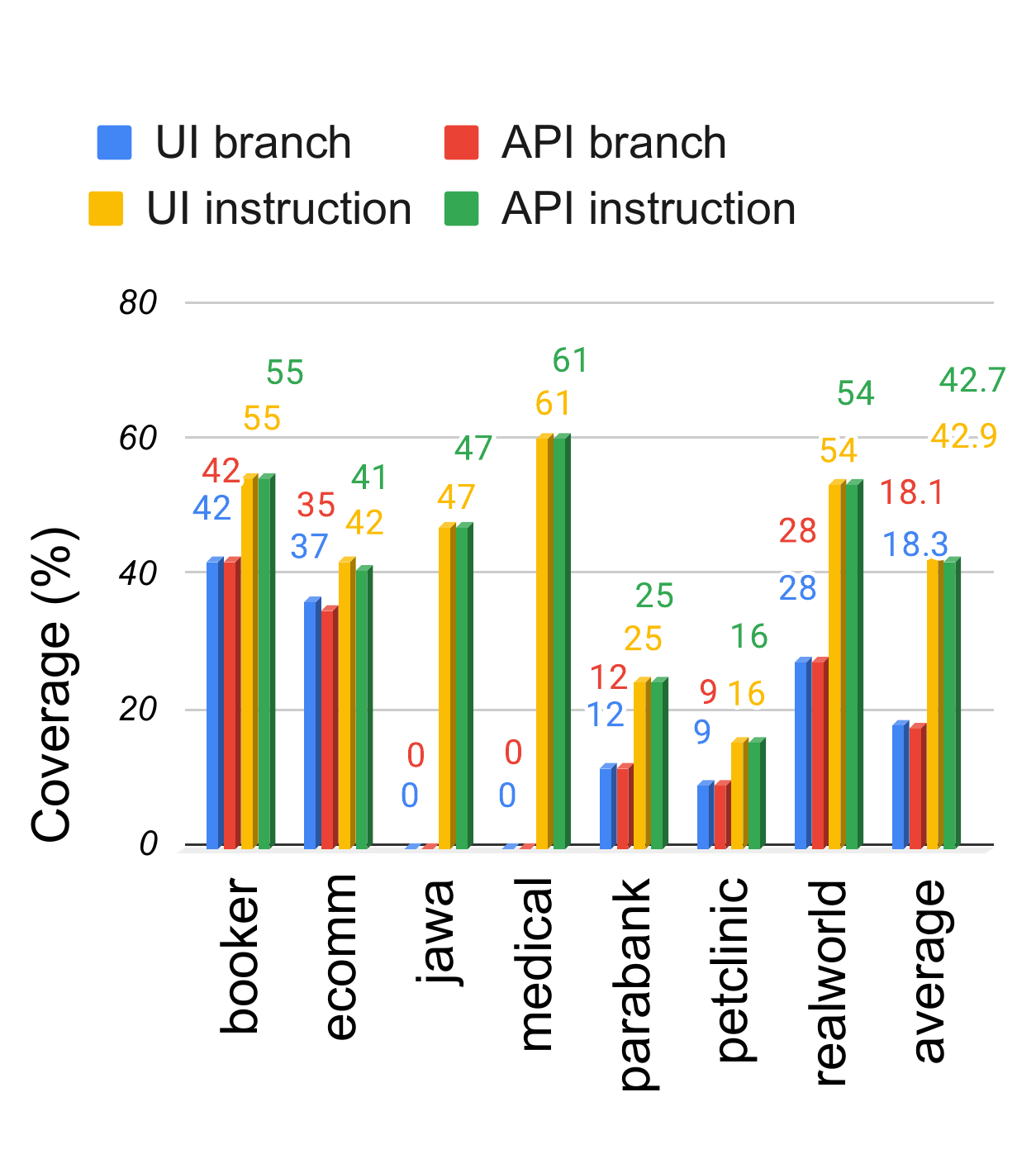}
\vspace*{-20pt}
\caption{\small Code coverage}
\label{fig:rq1cov}
\end{subfigure}
\begin{subfigure}[b]{0.44\columnwidth}
\includegraphics[trim=0.2cm 0.5cm 0.5cm 3.5cm, clip=true, width=\textwidth]{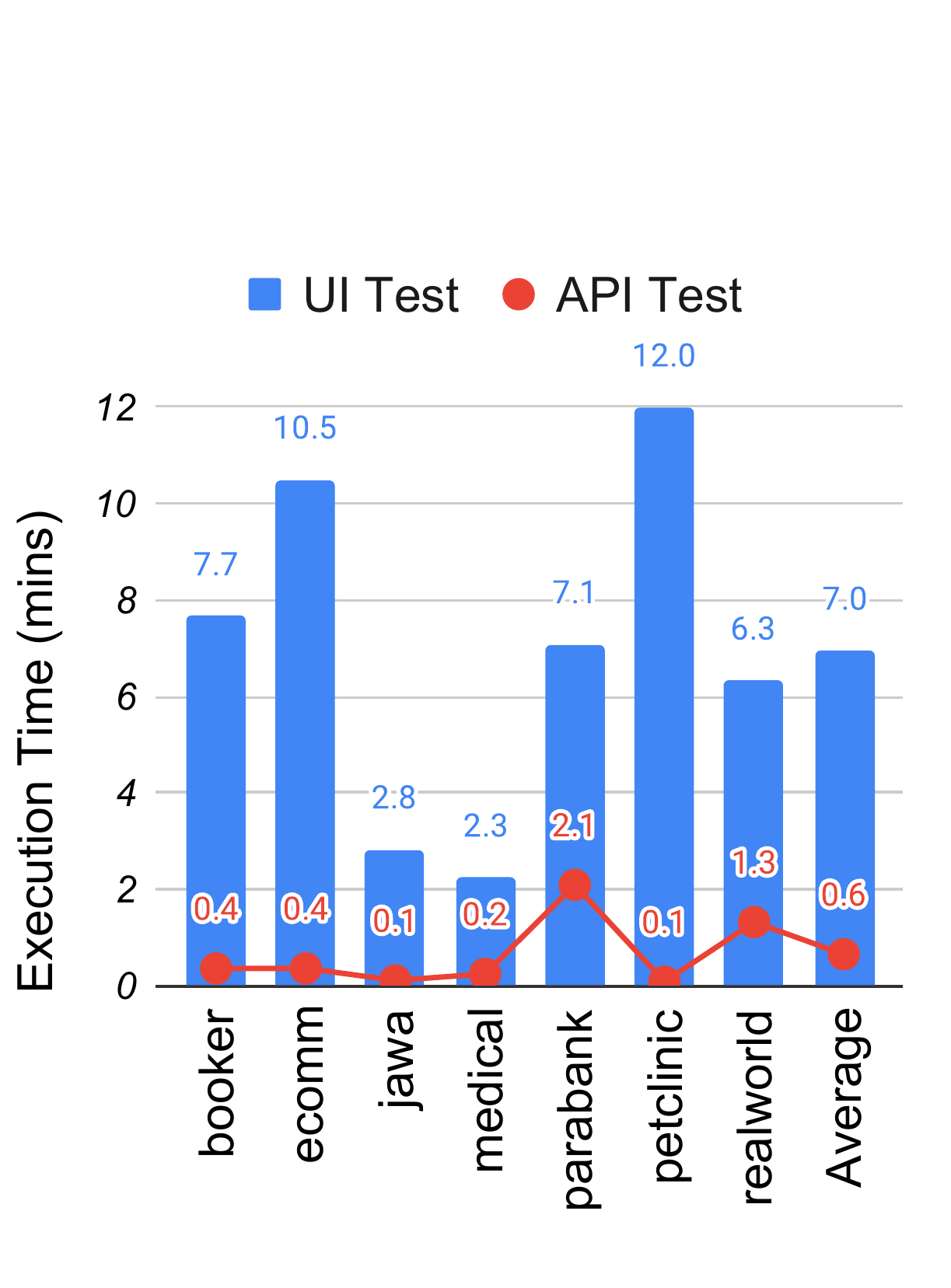}
\vspace*{-20pt}
\caption{\small Execution time}
\label{fig:rq1exec}
\end{subfigure}
\vspace*{-14pt}
\caption{Coverage rates and execution times of UI tests and carved API
  tests. 
  }
\label{fig:rq1}
\end{figure}

\subsection{\q{1}: Coverage Rates and Execution Efficiency of Tests}

~\autoref{fig:rq1cov} presents coverage rates for
UI tests and carved API tests. As the data illustrate, the coverage is identical
for all applications, except \texttt{\footnotesize ecomm}, for which instruction
and branch coverage of API test cases are marginally lower (by 1\% and 2\%
respectively). We suspect that this difference may have been due to API filtering. 
On
average, carved API test suites covered 18.1\% branches and 42.7\% instructions,
which is 0.2\% less than the coverage achieved by the UI test suites.  Thus,
overall, the carved API tests perform very well in matching the coverage rates
of UI test cases.

In terms of execution efficiency, however, there is a big difference between the
two types of test cases, as~\autoref{fig:rq1exec} shows.  On average, the UI
test suites took seven minutes to run, whereas the API test suites ran in about
0.6~minutes only---more than 10x improvement in execution efficiency. The
biggest improvement occurs for \texttt{\footnotesize petclinic}, for which the
UI test suite took 150 times longer to run than the API test suite. Even
with the smallest improvement, which occurs for \texttt{\footnotesize parabank},
the API tests executed over 3x faster than the UI tests (2.1~minutes versus
7.1~minutes, respectively).

\finding{The carved API tests match the coverage achieved by UI tests
  while executing significantly (10x) faster than UI tests. Thus, carved API
  tests can be employed for improving test execution efficiency, without
  incurring loss in coverage of server-side code.}

\subsection{\q{2}: Accuracy of Inferred OpenAPI Specification}

%

%


\header{Goals and Measures} To measure the effectiveness of \toolname in inferring API specifications, we compute precision, recall, and $F_1$ scores for the generated OpenAPI specification ($S_{gen}$) against the existing OpenAPI specification, considered the ground truth ($S_{gt}$), for each subject. We compute these scores for resource paths and operations (HTTP methods) defined on resource paths, and for the specification generated from the API graphs computed after carving and probing. Precision and recall are computed in the usual way, based on true positives, false positives, and false negatives. A path/operation is considered true positive if it occurs in both $S_{gen}$ and $S_{gt}$, false positive if it occurs in $S_{gen}$ but not in $S_{gt}$, and false negative if it occurs in $S_{gt}$ but not in $S_{gen}$. $F_1$ score is the harmonic mean of precision and recall.
In addition to these metrics, we measure \textit{duplication factor} for $S_{gen}$. A duplication occurs when multiple paths/operations in $S_{gen}$ correspond to the one path/operation in $S_{gt}$. We map paths in $S_{gen}$ to paths in $S_{gt}$, and compute duplication factor as (\# mapped paths in $S_{gt}$ / \# mapped paths in $S_{gen}$). The computed value ranges from 0~to 1, with higher values indicating less duplication (the value~1 means there is no repetition of API endpoints in $S_{gen}$). 
The presence of duplication causes $S_{gen}$ to contain redundant paths/operations that can be combined.



\begin{table}[]
\centering
\caption{\small Precision, recall, and F1 scores achieved for API specification inference.}
\vspace*{-5pt}  
\label{table:rq2_spec}
\renewcommand{\arraystretch}{1.05}
\resizebox{0.99\columnwidth}{!}{
\begin{tabular}{lrrrrrrrr}
\toprule
& \multicolumn{3}{c}{Path} & \multicolumn{5}{c}{Operation} \\
\cmidrule(lr){2-4}\cmidrule(lr){5-9}
Tool & Pr & \multicolumn{1}{c}{Re} & \multicolumn{1}{c}{$F_1$} & Pr & \multicolumn{1}{c}{Re} & \multicolumn{1}{c}{$F_1$} & \multicolumn{1}{c}{Pr*} & \multicolumn{1}{c}{$F_1$*} \\
\midrule
Carver        & 1.00   & 0.49   & 0.32   & 0.85 & 0.46 & 0.28 & 1.00  & 0.31  \\
Carver+Prober & 0.98   & 0.56   & 0.35   & 0.48 & 0.54 & 0.25 & 0.95  & 0.34 \\
\bottomrule
\end{tabular}
}
\vspace{1.5em}
\end{table}

\header{Results and Analysis} \autoref{table:rq2_spec} presents the precision, recall, and F1 scores for specification inference. In terms of resource paths, \toolname achieves 100\% precision for specifications created after carving and 98\% precision after probing (Column~2). The recall after the carving phase is 49\%, which the probing phase improves to 56\%---a gain of 14\% (Column~3). The probing phase is intended to address incompleteness in the API calls observed during UI navigation; the result shows that it achieves that to some degree and with only a small reduction in precision. The overall recall at 56\% is somewhat low, which is a consequence of the incompleteness inherent in dynamic analysis. This could be addressed via improvements in crawling or providing higher coverage UI test suites as input to \toolname. This concern is orthogonal to \toolname's core specification-inference and test-carving techniques.

In terms of operations, the results for recall (Column~6) after the carving phase is 46\%, which the probing phase improves to 54\%---a gain of 17\%. The precision value for operations is high (85\%) after carving, but there is a significant drop to 48\% after probing (Column~5). 
Upon closer inspection, we found that this is caused by operation probes, specifically the probes with  HTTP methods \code{OPTIONS} and \code{HEAD}; these requests are not handled correctly in any of the subjects. Ideally, an \code{OPTIONS} request should provide available operations for an API endpoint and the corresponding specification for the endpoint should contain \code{OPTIONS} as an operation. In all of our subjects, while the server returns a success
status (200~code) for an \code{OPTIONS} request, the corresponding operation is not documented in the specification. We consider this to be a specification
inconsistency with respect to application behavior; on ignoring this inconsistency, \toolname achieves 95\% precision for operations as well, shown as Pr* in~\autoref{table:rq2_spec} (Column~8). 


\finding{\toolname achieves high precision in inferring resource paths and operations. The probing phase of \toolname increases the recall and F1 scores, while not causing a significant reduction in precision.}

A manual analysis revealed that the path and operation precision drops from 1.0 to 0.98 and 0.95 because of one API endpoint found through probing in \texttt{\footnotesize realworld}. We verified that the resource path is indeed valid and provides a health-check for the service despite being absent in the specification, a potential inconsistency.
\autoref{table:rq2_dup} shows the operation inconsistencies that we found
per subject. The inconsistencies exposed by the carver are particularly
interesting because these \code{OPTIONS} and \code{HEAD} requests are actually
being used by the client---the application UI layer running in the browser---to communicate with the server. Recall that the carver uses only the requests captured during UI navigation. For example, the UI client of the \texttt{\footnotesize ecomm} application uses the  \code{OPTIONS} operation on 11~API
endpoints for server communication. These inconsistencies indicate room for potential improvements in the specifications, in particular, by documenting the \code{OPTIONS} HTTP method for API endpoints.

\begin{table}[]
\centering
\caption{\small Endpoints inferred, path/operation duplication found, and operation inconsistencies detected.}
\vspace*{-5pt}  
\label{table:rq2_dup}
\footnotesize
\renewcommand{\arraystretch}{1.05}
\resizebox{\columnwidth}{!}{
\begin{tabular}{l@{\hskip0.1em}r@{\hskip0.5em}r@{\hskip1.5em}r@{\hskip0.5em}r@{\hskip1em}r@{\hskip0.5em}r@{\hskip1em}r@{\hskip0.5em}r}
\toprule
          & \multicolumn{2}{c}{Endpoints} & \multicolumn{4}{c}{Duplication}                          & \multicolumn{2}{c}{Operation} \\
          \cmidrule(lr){4-7}
          & \multicolumn{2}{c}{Inferred}                                   & \multicolumn{2}{c}{Path} & \multicolumn{2}{c}{Operation} & \multicolumn{2}{c}{Inconsistencies}                                 \\
          \cmidrule(lr){2-3} \cmidrule(lr){4-5} \cmidrule(lr){6-7} \cmidrule(lr){8-9}
          & carver                    & car+pro                    & carver     & car+pro     & carver        & car+pro       & carver                   & car+pro                   \\
          \midrule
booker    & 8                         & 10                         & 0.89       & 0.91        & 0.92          & 0.94          & 0                        & 21                        \\
ecomm     & 13                        & 13                         & 0.81       & 0.81        & 0.82          & 0.82          & 11                       & 14                        \\
jawa      & 2                         & 2                          & 1.00       & 1.00        & 1.00          & 1.00          & 0                        & 2                         \\
medical   & 15                        & 16                         & 0.88       & 0.89        & 0.89          & 0.90          & 9                        & 17                        \\
parabank  & 9                         & 9                          & 1.00       & 1.00        & 1.00          & 1.00          & 0                        & 12                        \\
petclinic & 8                         & 12                         & 0.80       & 0.67        & 0.94          & 0.67          & 5                        & 18                        \\
realworld & 4                         & 5                          & 0.57       & 0.56        & 0.57          & 0.56          & 0                        & 18                        \\
\midrule
Average/Total   & 59                        & 67                         & 0.85       & 0.83        & 0.88          & 0.84          & 25                       & 102           \\
\bottomrule
\end{tabular}
}
\end{table}

\begin{figure*}[!t]
\centering
\begin{subfigure}[b]{0.49\textwidth}
\includegraphics[trim=0cm 0cm 0cm 0cm, clip=true, width=\textwidth
]{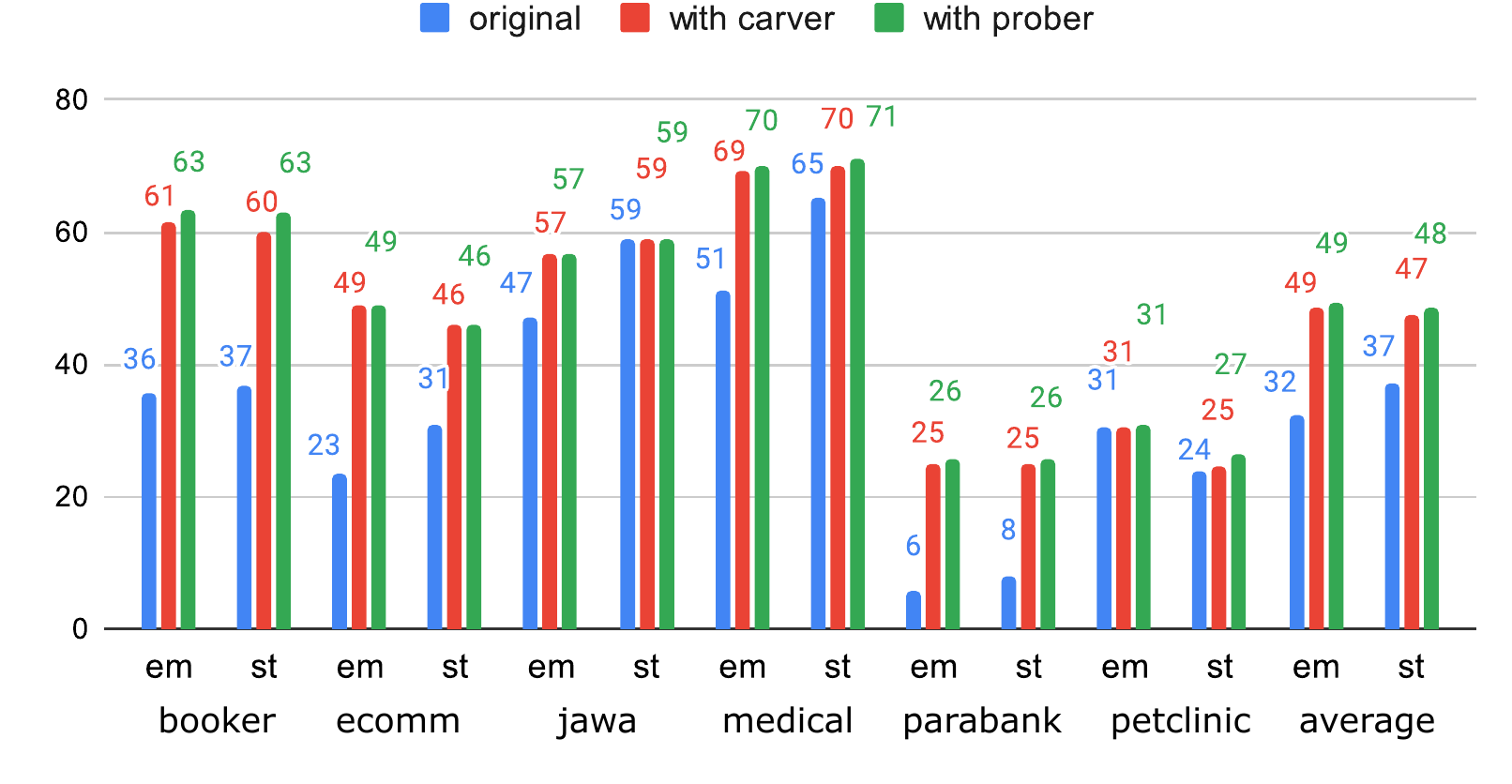}
\vspace*{-20pt}
\caption{\small Instruction coverage}
\label{fig:rq3_instr}
\end{subfigure}
\begin{subfigure}[b]{0.49\textwidth}
\centering
\includegraphics[trim=0cm 0cm 0cm 0cm, clip=true, width=\textwidth
]{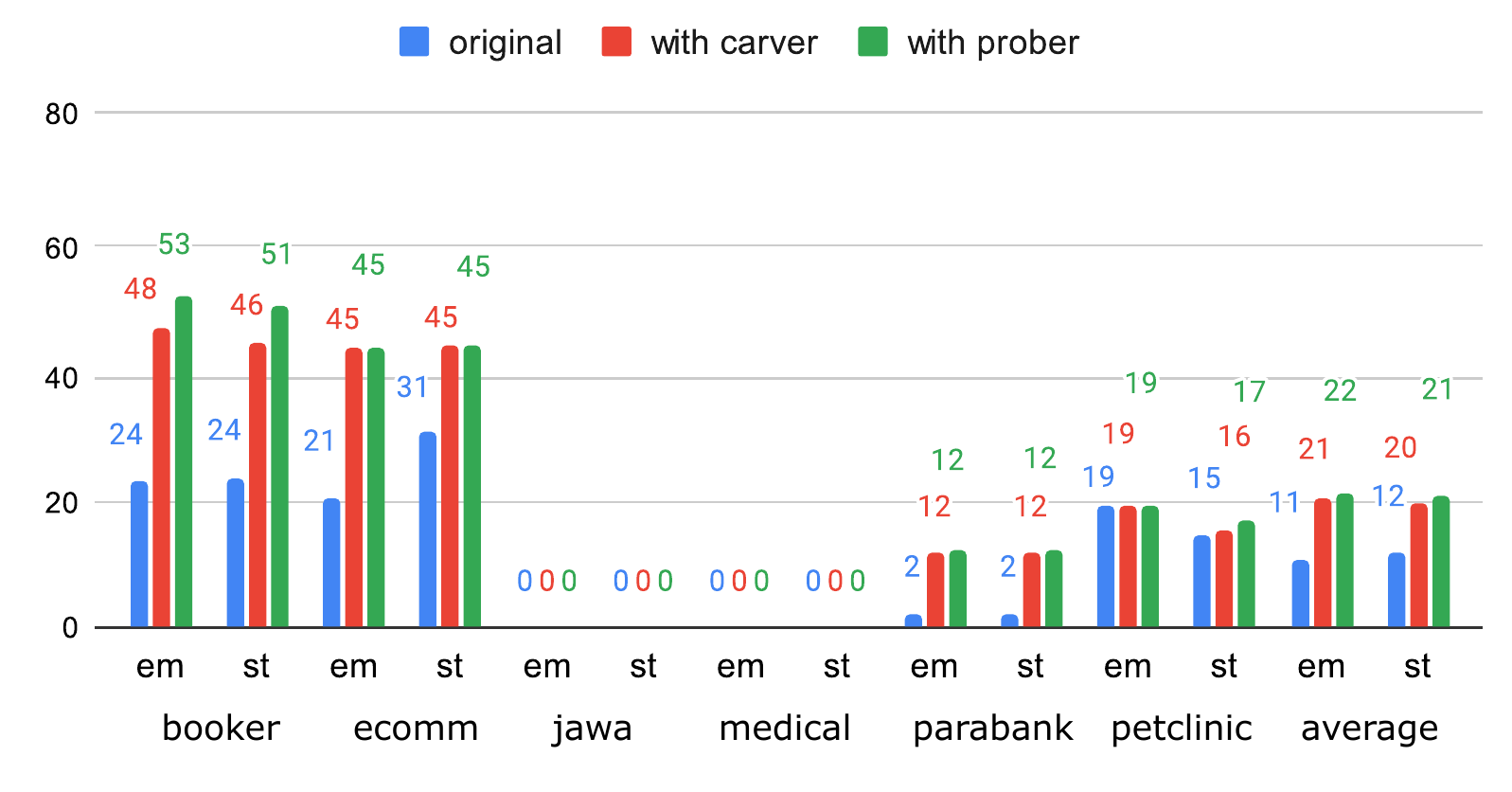}
\vspace*{-20pt}
\caption{\small Branch coverage}
\label{fig:rq3_branch}
\end{subfigure}
\vspace*{-4pt}
\caption{Augmentation effectiveness of carved tests: coverage rates of test suites generated by EvoMaster (em) and Schemathesis (st) before augmentation (original) and after augmentation with carved tests and probes.}
\vspace*{-12pt}
\label{fig:rq3_coverage}
\end{figure*}

Columns 4--7 of~\autoref{table:rq2_dup} show the duplication factor computed for the specification generated after the carving and probing phases. It can be seen that the duplication factor does not vary significantly for six of the subjects.
Path and operation duplication drops to 0.67 from 0.8 and 0.94, respectively, for \texttt{\footnotesize petclinic} because of the challenge in determining similarity of responses~(Algorithm~\ref{algo:specGen} lines 15--23). Recall that we use tree-based comparison to determine response similarity and,
in the case of petclinic, the server provides responses that are structurally dissimilar based on the back-end data differences. 

Endpoints covered (Columns 2--3 of~\autoref{table:rq2_dup}) improves for four of the seven subjects, with the largest increase of 50\% occurring for \texttt{\footnotesize petclinic}. Overall, endpoint coverage increases from 59 to 67, for 14\% increase, which is reflected in the path recall values~(Column 3 of \autoref{table:rq2_spec}) as well. 

\finding{\toolname can detect potential inconsistencies between API implementations and specifications, and could be leveraged for improving specifications.}

\subsection{\q{3}: Augmentation Effectiveness of Carved API Tests}

\header{Goals and Measures}
With \q{3}, we investigate the usefulness of carved API tests in enhancing the coverage rates achieved by EvoMaster~\cite{evomaster} and Schemathesis~\cite{schemathesis}. Specifically, we measure instruction and branch coverage of the test suites generated by those tools; then, we augment the test suites in two steps, by adding the carved API tests and the successful probes to the test suites, and measuring coverage gains in each augmentation step.


\header{Results and Analysis}
\autoref{fig:rq3_coverage} presents the results for \q{3}. It shows the instruction and branch coverage rates for the original API test suites generated by EvoMaster and Schemathesis and the two augmented test suites. Overall, our augmentation causes coverage increases in most instances, with a few exceptions (e.g., there are no instruction coverage gains for \texttt{\footnotesize jawa} with Schemathesis).
In terms of instructions, on average, the coverage of EvoMaster test suite increases from 32\% to 49\%, for a coverage gain of 52\%; for Schemathesis, coverage increases from 37\% to 48\%, for a coverage gain of 29\%. For branch coverage, augmentation has a bigger effect because of the low coverage rates of the original test suites. For EvoMaster, branch coverage gain is 99\%, increasing from 11\% to 22\%. For Schemathesis, branch coverage gain is 75\%, increasing from 12\% to 21\%. For both types of coverage, the gains for \texttt{\footnotesize booker}, \texttt{\footnotesize ecomm}, and \texttt{\footnotesize parabank} are substantial.

Moreover, additional coverage from probes, on top of the gains from carved tests, occurs in several instances. For example, the probes provide a considerable increase in branch coverage for \texttt{\footnotesize booker}---48\% to 53\% for EvoMaster and 46\% to 51\% for Schemathesis.



\finding{\toolname can significantly increase coverage achieved by EvoMaster and Schemathesis and, thus, can effectively complement such tools. The probing stage of \toolname can provide small additional gains on top of the gains from API tests carved from UI paths.
}

\begin{figure}[t]
  \begin{lstlisting}[language=Java
%      basicstyle=\tiny
    ]
/visits/{visitId}:          
  get:                 
    parameters:
      -name: visitId
       in: path
\end{lstlisting}
\vspace{-6pt}
\begin{lstlisting}[language=Java,
%      basicstyle=\tiny
    ]                
public ResponseEntity getVisit(Integer visitId) {
    Visit visit = this.clinicService.findVisitById(visitId);
    if (visit == null) { 
        /* Covered by ApiCarv, Schemathesis and Evomaster */
        return new ResponseEntity<>(HttpStatus.NOT_FOUND);
    }
    /* Covered by ApiCarv through probing */
    return new ResponseEntity<>(visitMapper.toVisitDto(visit), HttpStatus.OK);
}
\end{lstlisting}
\vspace{-9pt}
\caption{Example of an endpoint and the associated service code (from \code{petclinic}) that requires specific test data.}
\label{listing:testData}
\end{figure}

\begin{figure}[t]
  \begin{lstlisting}[language=Java
%      basicstyle=\tiny,
    ]
/user/signUp:
  post:
    params: {in: body, schema: {"user": string, "pass": string}} 
    responses: {200: {"status": string}}
      
/user/signIn:          
  post:   
    params: {in: body, schema: {"user": string, "pass": string}} 
    responses: {200:{"status": string, "token": string}} 
      
/cart/add:
  post:
    params: {name: token, in : query, schema : string}
\end{lstlisting}
\vspace{-9pt}
\caption{Example (from the \code{ecomm} application) of dependencies between API endpoints.}
\label{listing:apiDependencies}
\end{figure}

To understand how carved API test suites can complement API testing techniques, we performed an in-depth analysis of the differences in code coverage achieved by the testing tools and \toolname. Our analysis revealed two interesting high-level scenarios where carving API test suites from end-to-end UI test suites could improve the overall effectiveness of API-level testing of web applications. 

First, generating appropriate test data to cover API endpoints is a challenging aspect of API fuzzing, and carved API test suites can be leveraged to cover certain endpoints that require specific test data. An example of such a scenario from \code{petclinic} is shown in~\autoref{listing:testData} where covering the \code{/visits/visitId} endpoint requires providing a value for \code{visitId} that is already present in the application database. In this instance, \toolname leveraged the analysis of API calls observed during the carving phase~(derived from the UI test suite) to find a value of \code{visitId} that covers line~8 of method \code{getVisit()} and elicits a successful response from the service (\code{HttpStatus.OK}); EvoMaster and Schemathesis were unable to craft a request with a valid value for \code{visitId}.

Second, some API endpoints can have dependencies on other API endpoints. Consider the three API endpoints from the \code{ecomm} application shown in~\autoref{listing:apiDependencies}. The endpoint \code{/cart/add} requires a query parameter, \code{token}, that is generated by the server in response to a call to the \code{/user/signIn} endpoint. But, to invoke \code{/user/signIn}, a user must first be registered via the \code{/user/signUp} endpoint. EvoMaster could invoke \code{/user/signUp} successfully, but it could not execute the subsequent operations to sign in and add item to cart; Schemathesis could not cover any of these operations. 
In contrast, the API test suite carved from the UI test suite of \code{ecomm} could create a successful \code{POST\,/cart/add} request by satisfying these dependencies. The API calls in the carved test suite are constructed by the web application UI, which is developed to adhere to the API specification and valid API invocation sequences. Thus, such dependencies are inherently followed in sequences of API calls made along UI paths, and \toolname captures the call sequences by navigating those paths. 
The API testing tools, although they attempt to discover meaningful or valid API call sequences, could not generate requests that satisfy the endpoint dependencies in this instance.

Modern web applications are commonly tested in the industry using end-to-end UI test suites, which are often manually created. It is efficient and convenient to develop such UI test suites because of the mature UI testing eco-system and the fact that application business logic is easily translated into a sequence of UI actions on the web interface. API tests carved from such UI test suites can, therefore, contain realistic test data and encapsulate the application business logic in a sequence of API calls. Given these characteristics, carved test suites could effectively complement the API tests generated by API testing tools, such as EvoMaster and Schemathesis.

\subsection{Threats to Validity}
Our study may suffer external and internal threats to validity. In terms of external threats, we used seven web applications and two REST API testing tools. Our selection of web applications was constrained to applications that implement RESTful services and have OpenAPI specifications available to serve as ground truth; also, our requirement of measuring code coverage on APIs further constrained the candidate applications. Future evaluation with more and varied web applications will help confirm whether our results generalize. Our selection of REST API testing tools was guided by a recent study~\cite{kim:2022} that showed EvoMaster and Schemathesis to be the most effective tools, in terms of coverage achieved, among the studied black-box testing tools. Another threat is the use of existing OpenAPI specifications as ground truth. We think this is a reasonable choice for our experiments and we use them with the expectation that they may have some inconsistencies.
As for internal threats, there may be bugs in \toolname and our data-collection scripts. We mitigated these threats by implementing automated unit test cases for \toolname and manually checking random samples of our results. We also make \toolname and our experiment artifacts available~\cite{toollink} to enable replication of our results.


\section{Related Work}
\label{sec:related}
To the best of our knowledge, our work is the first to propose carving of API-level tests from UI-level test executions.
Several papers have explored carving of unit-level tests from system-level executions via code instrumentation (e.g.,~\cite{DBLP:journals/tse/ElbaumCDJ09, ICSM-2007-JoshiO, DBLP:conf/sigsoft/XuRTQ07}). Elbaum et al.~\cite{DBLP:journals/tse/ElbaumCDJ09} present a technique for carving unit-level tests from system tests consisting of Java-based code exercising the application end-to-end. 
Other techniques~\cite{ICSM-2007-JoshiO} selectively capture and replay events and interactions
between selected program components and the rest of the application, using simplified state representations or they aim~\cite{DBLP:conf/sigsoft/XuRTQ07} at enhancing replay efficiency 
by mixing action-based and state-based checkpointing. These approaches highlight different advantages of carved tests compared to the original tests, 
such as their execution efficiency and robustness to program changes. Carved unit tests are shown~\cite{DBLP:journals/tse/ElbaumCDJ09} to be orders of magnitudes faster than the original executions, while retaining most of their fault-detection capabilities. These benefits also motivate our work, and our evaluation demonstrates the significantly superior execution efficiency of carved API tests compared with UI-level tests, with negligible loss in code coverage. 

Dynamic specification mining has mostly focused on mining behavioral models of a program from its execution traces (e.g.,~\cite{10.1145/503272.503275,ICSE-2008-LorenzoliMP, DBLP:conf/icsm/PradelBG10}). 
These models capture relations between data values and component interactions, to allow for accurate analysis and verification of the software. More relevant to our work are the
approaches recently suggested for mining OpenAPI specifications. Several works propose inferring OpenAPI specifications from web API documentation pages. AutoREST~\cite{DBLP:conf/icsoc/CaoFB17}  
 infers API specifications from HTML-based documentation via selection of web pages that likely contain information relevant to the 
specification. It applies a set of rules to extract relevant information from the pages and construct the specification. 
D2Spec~\cite{8595229} uses machine-learning techniques to extract the base URL, path templates, and HTTP methods from crawled documentation pages.
A different approach that uses dynamic information is taken in~\cite{10.1007/978-3-319-61482-3_16}, which generates web API specifications from example request-response pairs. 
Closest work to ours is SpyREST~\cite{7372015}, which intercepts 
HTTP requests and applies a simple heuristic for identifying path parameters, by considering numeric path items 
and using regular expression matching. In contrast, our approach infers path parameters via API-graph analysis and API probing. We tried to execute SpyREST for comparison with 
\toolname's specification inference, but its service failed to work.

\section{Conclusion and Future Work}

We presented \toolname{}, a first-of-its-kind technique and tool for carving API tests and specifications from UI tests. Our evaluation on seven open-source web applications showed that (1) carved API tests achieve similar coverage as the UI tests that they are created from, but with significantly less (10x) execution time, (2) \toolname{} achieves high precision in inferring API specifications, and (3) \toolname{} can increase the coverage achieved by automated API test generators.

There are several directions in which our approach could be extended in future work, including development of techniques for improving the inferred specifications to take them closer to developer-written specifications, enhancing the specifications with information (e.g., example values) that can be leveraged by automated REST API test generators, and improving the recall of specification inference via novel crawling techniques aimed at discovering the server-side APIs of a web application.



\balance
\bibliographystyle{IEEEtran}
\bibliography{paper}

\begin{thebibliography}{10}
\providecommand{\url}[1]{#1}
\csname url@samestyle\endcsname
\providecommand{\newblock}{\relax}
\providecommand{\bibinfo}[2]{#2}
\providecommand{\BIBentrySTDinterwordspacing}{\spaceskip=0pt\relax}
\providecommand{\BIBentryALTinterwordstretchfactor}{4}
\providecommand{\BIBentryALTinterwordspacing}{\spaceskip=\fontdimen2\font plus
\BIBentryALTinterwordstretchfactor\fontdimen3\font minus
  \fontdimen4\font\relax}
\providecommand{\BIBforeignlanguage}[2]{{%
\expandafter\ifx\csname l@#1\endcsname\relax
\typeout{** WARNING: IEEEtran.bst: No hyphenation pattern has been}%
\typeout{** loaded for the language `#1'. Using the pattern for}%
\typeout{** the default language instead.}%
\else
\language=\csname l@#1\endcsname
\fi
#2}}
\providecommand{\BIBdecl}{\relax}
\BIBdecl

\bibitem{fielding2000architectural}
R.~T. Fielding, \emph{Architectural styles and the design of network-based
  software architectures}.\hskip 1em plus 0.5em minus 0.4em\relax University of
  California, Irvine Irvine, 2000, vol.~7.

\bibitem{openapi}
{Open API Initiative}, ``Openapi specification,''
  \url{https://spec.openapis.org/oas/latest.html}, 2022, accessed: 2022-01-01.

\bibitem{apiblueprint}
\BIBentryALTinterwordspacing
``{API Blueprint},'' 2022, accessed: Sep 1, 2022. [Online]. Available:
  \url{https://apiblueprint.org/}
\BIBentrySTDinterwordspacing

\bibitem{raml}
\BIBentryALTinterwordspacing
``{RAML},'' 2022, accessed: Sep 1, 2022. [Online]. Available:
  \url{https://raml.org/}
\BIBentrySTDinterwordspacing

\bibitem{martinlopez2019criteria}
A.~Martin-Lopez, S.~Segura, and A.~Ruiz-Cort\'{e}s, ``{Test Coverage Criteria
  for RESTful Web APIs},'' in \emph{Proceedings of the 10th ACM SIGSOFT
  International Workshop on Automating TEST Case Design, Selection, and
  Evaluation}, 2019, p. 15–21.

\bibitem{arcuri2019restful}
A.~Arcuri, ``Restful api automated test case generation with evomaster,''
  \emph{ACM Transactions on Software Engineering and Methodology (TOSEM)},
  vol.~28, no.~1, pp. 1--37, 2019.

\bibitem{atlidakis2019restler}
V.~Atlidakis, P.~Godefroid, and M.~Polishchuk, ``Restler: Stateful rest api
  fuzzing,'' in \emph{2019 IEEE/ACM 41st International Conference on Software
  Engineering (ICSE)}.\hskip 1em plus 0.5em minus 0.4em\relax Montreal, QC,
  Canada: IEEE, 2019, pp. 748--758.

\bibitem{viglianisi2020resttestgen}
E.~Viglianisi, M.~Dallago, and M.~Ceccato, ``Resttestgen: automated black-box
  testing of restful apis,'' in \emph{2020 IEEE 13th International Conference
  on Software Testing, Validation and Verification (ICST)}.\hskip 1em plus
  0.5em minus 0.4em\relax IEEE, 2020, pp. 142--152.

\bibitem{karlsson2020quickrest}
S.~Karlsson, A.~{\v{C}}au{\v{s}}evi{\'c}, and D.~Sundmark, ``Quickrest:
  Property-based test generation of openapi-described restful apis,'' in
  \emph{13th International Conference on Software Testing, Validation and
  Verification (ICST)}.\hskip 1em plus 0.5em minus 0.4em\relax IEEE, 2020, pp.
  131--141.

\bibitem{martin2020restest}
A.~Martin-Lopez, S.~Segura, and A.~Ruiz-Cort{\'e}s, ``Restest: Black-box
  constraint-based testing of restful web apis,'' in \emph{International
  Conference on Service-Oriented Computing}.\hskip 1em plus 0.5em minus
  0.4em\relax Springer, 2020, pp. 459--475.

\bibitem{godefroid2020intelligent}
P.~Godefroid, B.-Y. Huang, and M.~Polishchuk, ``Intelligent rest api data
  fuzzing,'' in \emph{Proceedings of the 28th ACM Joint Meeting on European
  Software Engineering Conference and Symposium on the Foundations of Software
  Engineering}, 2020, pp. 725--736.

\bibitem{Zac2021schemathesis}
\BIBentryALTinterwordspacing
Z.~Hatfield-Dodds and D.~Dygalo, ``Deriving semantics-aware fuzzers from web
  api schemas,'' 2021. [Online]. Available:
  \url{https://arxiv.org/abs/2112.10328}
\BIBentrySTDinterwordspacing

\bibitem{corradiniautomated}
D.~Corradini, A.~Zampieri, M.~Pasqua, E.~Viglianisi, M.~Dallago, and
  M.~Ceccato, ``Automated black-box testing of nominal and error scenarios in
  restful apis,'' \emph{Software Testing, Verification and Reliability}, p.
  e1808, 2022.

\bibitem{segura2017metamorphic}
S.~Segura, J.~A. Parejo, J.~Troya, and A.~Ruiz-Cort{\'e}s, ``Metamorphic
  testing of restful web apis,'' \emph{IEEE Transactions on Software
  Engineering (TSE)}, pp. 1083--1099, 2017.

\bibitem{laranjeiro2021black}
N.~Laranjeiro, J.~Agnelo, and J.~Bernardino, ``A black box tool for robustness
  testing of rest services,'' \emph{IEEE Access}, pp. 24\,738--24\,754, 2021.

\bibitem{stallenberg2021improving}
D.~Stallenberg, M.~Olsthoorn, and A.~Panichella, ``Improving test case
  generation for rest apis through hierarchical clustering,'' in \emph{2021
  36th IEEE/ACM International Conference on Automated Software Engineering
  (ASE)}.\hskip 1em plus 0.5em minus 0.4em\relax IEEE, 2021, pp. 117--128.

\bibitem{wu2022combinatorial}
H.~Wu, L.~Xu, X.~Niu, and C.~Nie, ``Combinatorial testing of restful apis,'' in
  \emph{ACM/IEEE International Conference on Software Engineering (ICSE)},
  2022.

\bibitem{liu2022morest}
Y.~Liu, Y.~Li, G.~Deng, Y.~Liu, R.~Wan, R.~Wu, D.~Ji, S.~Xu, and M.~Bao,
  ``Morest: Model-based restful api testing with execution feedback,''
  \emph{arXiv preprint arXiv:2204.12148}, 2022.

\bibitem{10.1145/3491038}
B.~Marculescu, M.~Zhang, and A.~Arcuri, ``On the faults found in rest apis by
  automated test generation,'' \emph{ACM Trans. Softw. Eng. Methodol.},
  vol.~31, no.~3, 2022.

\bibitem{springfox}
\BIBentryALTinterwordspacing
``{SpringFox: Automated JSON API documentation for API's built with Spring},''
  2022, accessed: Sep 1, 2022. [Online]. Available:
  \url{https://springfox.github.io/springfox/}
\BIBentrySTDinterwordspacing

\bibitem{springdoc}
\BIBentryALTinterwordspacing
``springdoc-openapi,'' 2022, accessed: Sep 1, 2022. [Online]. Available:
  \url{https://springdoc.org/}
\BIBentrySTDinterwordspacing

\bibitem{springboot}
\BIBentryALTinterwordspacing
``{SpringBoot},'' 2022, accessed: Sep 1, 2022. [Online]. Available:
  \url{https://spring.io/projects/spring-boot/}
\BIBentrySTDinterwordspacing

\bibitem{DBLP:journals/tse/ElbaumCDJ09}
S.~G. Elbaum, H.~N. Chin, M.~B. Dwyer, and M.~Jorde, ``Carving and replaying
  differential unit test cases from system test cases,'' \emph{{IEEE} Trans.
  Software Eng.}, vol.~35, no.~1, pp. 29--45, 2009.

\bibitem{ICSM-2007-JoshiO}
S.~Joshi and A.~Orso, ``{SCARPE: A Technique and Tool for Selective Capture and
  Replay of Program Executions},'' in \emph{{Proceedings of the 23rd
  International Conference on Software Maintenance}}.\hskip 1em plus 0.5em
  minus 0.4em\relax {IEEE}, 2007, pp. 234--243.

\bibitem{DBLP:conf/sigsoft/XuRTQ07}
G.~Xu, A.~Rountev, Y.~Tang, and F.~Qin, ``Efficient checkpointing of java
  software using context-sensitive capture and replay,'' in
  \emph{{ESEC/SIGSOFT} {FSE}}.\hskip 1em plus 0.5em minus 0.4em\relax {ACM},
  2007, pp. 85--94.

\bibitem{evomaster}
\BIBentryALTinterwordspacing
``{EvoMaster: A Tool For Automatically Generating System-Level Test Cases},''
  2022, accessed: Sep 1, 2022. [Online]. Available:
  \url{https://github.com/EMResearch/EvoMaster}
\BIBentrySTDinterwordspacing

\bibitem{schemathesis}
\BIBentryALTinterwordspacing
``schemathesis,'' 2022, accessed: Sep 1, 2022. [Online]. Available:
  \url{https://github.com/schemathesis/schemathesis}
\BIBentrySTDinterwordspacing

\bibitem{zac2022schemathesis}
Z.~Hatfield-Dodds and D.~Dygalo, ``Deriving semantics-aware fuzzers from web
  api schemas,'' in \emph{2022 IEEE/ACM 44th International Conference on
  Software Engineering: Companion Proceedings (ICSE-Companion)}, 2022, pp.
  345--346.

\bibitem{toollink}
``{Carving UI tests suites to generate API tests and API specification},''
  \url{https://github.com/apicarve/apicarver}, 2022.

\bibitem{realworld}
{Gérôme Grignon, Manuel Vila}, ``The mother of all demo apps,''
  \url{https://github.com/gothinkster/realworld}, 2022, accessed: 2022-01-01.

\bibitem{uriTemplate}
\BIBentryALTinterwordspacing
J.~Gregorio, R.~Fielding, M.~Hadley, M.~Nottingham, , and D.~Orchard, ``{URI
  Template},'' RFC 6570, 2012. [Online]. Available:
  \url{https://www.rfc-editor.org/info/rfc6570}
\BIBentrySTDinterwordspacing

\bibitem{Grechanik:2009:MEG:1555001.1555055}
M.~Grechanik, Q.~Xie, and C.~Fu, ``Maintaining and evolving {GUI-directed} test
  scripts,'' in \emph{Proceedings of 31st International Conference on Software
  Engineering}, ser. ICSE 2009.\hskip 1em plus 0.5em minus 0.4em\relax IEEE,
  2009, pp. 408--418.

\bibitem{apted}
M.~Pawlik and N.~Augsten, ``Tree edit distance: Robust and memory-efficient,''
  \emph{Inf. Syst.}, vol.~56, pp. 157--173, 2016.

\bibitem{crawljax}
A.~Mesbah, A.~van Deursen, and S.~Lenselink, ``Crawling ajax-based web
  applications through dynamic analysis of user interface state changes,''
  \emph{ACM Transactions on the Web}, vol.~6, no.~1, pp. 3:1--3:30, 2012.

\bibitem{fraggen}
R.~K. Yandrapally and A.~Mesbah, ``Fragment-based test generation for web
  apps,'' \emph{IEEE Transactions on Software Engineering}, p. 16 pages, 2022.

\bibitem{cdp}
{Google}, ``Chrome devtools protocol,''
  \url{https://chromedevtools.github.io/devtools-protocol/}, 2022, accessed:
  2022-01-01.

\bibitem{selenium}
``{Selenium} web browser automation,'' \url{https://www.selenium.dev/}, 2022,
  accessed: 2022-07-01.

\bibitem{kim:2022}
\BIBentryALTinterwordspacing
M.~Kim, Q.~Xin, S.~Sinha, and A.~Orso, ``Automated test generation for rest
  apis: No time to rest yet,'' in \emph{Proceedings of the 31st ACM SIGSOFT
  International Symposium on Software Testing and Analysis}.\hskip 1em plus
  0.5em minus 0.4em\relax Association for Computing Machinery, 2022, p.
  289–301. [Online]. Available: \url{https://doi.org/10.1145/3533767.3534401}
\BIBentrySTDinterwordspacing

\bibitem{jacoco}
\BIBentryALTinterwordspacing
``{JaCoCo Java Code Coverage Library},'' 2022, accessed: Sep 1, 2022. [Online].
  Available: \url{https://www.eclemma.org/jacoco/}
\BIBentrySTDinterwordspacing

\bibitem{istanbul}
{Apache}, ``https://istanbul.js.org/,'' \url{https://istanbul.js.org/}, 2022,
  accessed: 2022-09-01.

\bibitem{10.1145/503272.503275}
G.~Ammons, R.~Bod\'{\i}k, and J.~R. Larus, ``Mining specifications,'' ser. POPL
  '02.\hskip 1em plus 0.5em minus 0.4em\relax ACM, 2002, p. 4–16.

\bibitem{ICSE-2008-LorenzoliMP}
D.~Lorenzoli, L.~Mariani, and M.~Pezzè, ``{Automatic generation of software
  behavioral models},'' in \emph{{Proceedings of the 30th International
  Conference on Software Engineering}}, W.~Schäfer, M.~B. Dwyer, and V.~Gruhn,
  Eds.\hskip 1em plus 0.5em minus 0.4em\relax {ACM}, 2008, pp. 501--510.

\bibitem{DBLP:conf/icsm/PradelBG10}
M.~Pradel, P.~Bichsel, and T.~R. Gross, ``A framework for the evaluation of
  specification miners based on finite state machines,'' in
  \emph{{ICSM}}.\hskip 1em plus 0.5em minus 0.4em\relax {IEEE} Computer
  Society, 2010, pp. 1--10.

\bibitem{DBLP:conf/icsoc/CaoFB17}
H.~Cao, J.~Falleri, and X.~Blanc, ``Automated generation of {REST} {API}
  specification from plain {HTML} documentation,'' in \emph{{ICSOC}}, ser.
  Lecture Notes in Computer Science, vol. 10601.\hskip 1em plus 0.5em minus
  0.4em\relax Springer, 2017, pp. 453--461.

\bibitem{8595229}
J.~Yang, E.~Wittern, A.~T. Ying, J.~Dolby, and L.~Tan, ``Towards extracting web
  api specifications from documentation,'' in \emph{2018 IEEE/ACM 15th
  International Conference on Mining Software Repositories (MSR)}, 2018, pp.
  454--464.

\bibitem{10.1007/978-3-319-61482-3_16}
H.~Ed-douibi, J.~L. C{\'a}novas~Izquierdo, and J.~Cabot, ``Example-driven web
  api specification discovery,'' in \emph{Modelling Foundations and
  Applications}, A.~Anjorin and H.~Espinoza, Eds.\hskip 1em plus 0.5em minus
  0.4em\relax Springer International Publishing, 2017, pp. 267--284.

\bibitem{7372015}
S.~M. Sohan, C.~Anslow, and F.~Maurer, ``Spyrest: Automated restful api
  documentation using an http proxy server (n),'' in \emph{2015 30th IEEE/ACM
  International Conference on Automated Software Engineering (ASE)}, 2015, pp.
  271--276.

\end{thebibliography}

\end{document}